\documentclass[conference]{IEEEtran}
\IEEEoverridecommandlockouts

\usepackage{cite}
\usepackage{amsmath,amssymb,amsfonts}
\usepackage{algorithmic}
\usepackage{graphicx}
\usepackage{textcomp}
\usepackage{xcolor}

\usepackage{algorithm}
\usepackage{color}
\usepackage{booktabs}
\usepackage{subcaption}
\usepackage[section]{placeins}
\usepackage[english]{babel}
\usepackage{breakurl}
\usepackage[hyphens]{url}

\def\BibTeX{{\rm B\kern-.05em{\sc i\kern-.025em b}\kern-.08em
    T\kern-.1667em\lower.7ex\hbox{E}\kern-.125emX}}

\definecolor{dkred}{rgb}{.6,0,0}
\definecolor{dkgreen}{rgb}{0,.5,0}

\usepackage{listings}
\usepackage{newfloat}
\floatstyle{ruled}
\newfloat{listing}{tb}{lst}{}
\floatname{listing}{Listing}

\makeatletter
\newcommand{\linebreakand}{%
\end{@IEEEauthorhalign}
\hfill\mbox{}\par
\mbox{}\hfill\begin{@IEEEauthorhalign}
}
\makeatother

\newif\ifdraft\draftfalse

\newcommand{\rebut}[1]{#1}

\newcommand{\olremark}[2]{{\color{dkgreen}(#1: #2)}}
\newcommand{\varemark}[2]{{\color{dkred}(#1: #2)}}
\newcommand{\smremark}[2]{{\color{red}(#1: #2)}}
\newcommand{\eiremark}[2]{{\color{cyan}(#1: #2)}}
\newcommand{\ftremark}[2]{{\color{teal}(#1: #2)}}

\newcommand{\lnegremark}[2]{{\color{cyan}(#1: #2)}}
\newcommand{\loliremark}[2]{{\color{orange}(#1: #2)}}
\ifdraft
\newcommand{\va}[1]{\varemark{VA}{#1}}
\newcommand{\gd}[1]{\olremark{GD}{#1}}
\newcommand{\mm}[1]{\ftremark{MM}{#1}}
\newcommand{\lneg}[1]{\lnegremark{LN}{#1}}
\newcommand{\loli}[1]{\loliremark{LO}{#1}}
\newcommand{\ez}[1]{\varemark{EZ}{#1}}
\newcommand{\sm}[1]{\smremark{SM}{#1}}
\newcommand{\ei}[1]{\eiremark{EI}{#1}}
\else
\newcommand{\va}[1]{}
\newcommand{\gd}[1]{}
\newcommand{\mm}[1]{}
\newcommand{\lneg}[1]{}
\newcommand{\loli}[1]{}
\newcommand{\ez}[1]{}
\newcommand{\sm}[1]{}
\newcommand{\ei}[1]{}
\fi

\renewcommand\subsubsection[1]{\vspace{5pt}\noindent \textbf{#1.}}

\title{Helping LLMs Improve Code Generation Using Feedback from Testing and Static Analysis}

\author{
	\IEEEauthorblockN{Greta Dolcetti\IEEEauthorrefmark{1}, Vincenzo Arceri\IEEEauthorrefmark{2}, Eleonora Iotti\IEEEauthorrefmark{2}, Sergio Maffeis\IEEEauthorrefmark{3}, Agostino Cortesi\IEEEauthorrefmark{1}, and Enea Zaffanella\IEEEauthorrefmark{2}}
	\IEEEauthorblockA{\IEEEauthorrefmark{1}Ca' Foscari University  of Venice, Venice, Italy\\
		Email: \{greta.dolcetti, cortesi\}@unive.it}
	\IEEEauthorblockA{\IEEEauthorrefmark{2}University of Parma, Parma, Italy\\
		Email: \{vincenzo.arceri, eleonora.iotti, enea.zaffanella\}@unipr.it}
	\IEEEauthorblockA{\IEEEauthorrefmark{3}Imperial College London, London, United Kingdom\\
		Email: sergio.maffeis@imperial.ac.uk}
}

\usepackage{enumitem}

\begin{document}
\maketitle

\begin{abstract}
    Large Language Models (LLMs) are one of the most promising
    developments in the field of artificial intelligence, and the software
    engineering community has readily noticed their potential role in the software development life-cycle.
    Developers routinely ask LLMs to generate code snippets, increasing productivity but also potentially introducing ownership, privacy, correctness, and security issues.
    
    Previous work highlighted how code generated by mainstream commercial LLMs is often not safe, containing vulnerabilities, bugs, and code smells. 
    In this paper, we present a framework that leverages testing and static analysis to assess the quality, and guide the self-improvement, of code generated by general-purpose, open-source LLMs.
    
    First, we ask LLMs to generate C code to solve a number of programming tasks. 
    Then we employ ground-truth tests to assess the (in)correctness of the generated code, and a static analysis tool to detect potential safety vulnerabilities.
    Next, we assess the models ability to evaluate the generated code, by asking them to detect errors and vulnerabilities. 
    Finally, we test the models ability to fix the generated code, providing the reports produced during the static analysis and incorrectness evaluation phases as feedback.
    
    \rebut{Our results show that models often produce incorrect code, and that the generated code can include safety issues.} 
    Moreover, they perform very poorly at detecting either issue.
    On the positive side, we observe a substantial ability to fix flawed code when provided with information about failed tests or potential vulnerabilities, indicating a promising avenue for improving the safety of LLM-based code generation tools.
    
\end{abstract}

\begin{IEEEkeywords}
    Large Language Models, Code Generation, Static Analysis, Code Repair
\end{IEEEkeywords}

\section{Introduction}\label{sec:intro}
The impressive acceleration we are witnessing in the offer of new and
increasingly efficient Large Language Models (LLMs), on the one hand fascinates
many researchers and on the other creates distrust and a priori attitudes
of refusal.

One of the most worrying aspects is the ability of these
models to generate source code automatically, with the perceived risk
of losing control of the reliability of software systems, in a context in which
all the main aspects of our lives depend on them. Many studies
already aim at demonstrating the intrinsic weaknesses
of LLMs and their applications~\cite{abs-2401-05949}. 
A comprehensive list of such weaknesses is reported by OWASP (Open Web Application
Security Project) in its Top 10 vulnerabilities for
LLMs,\footnote{\url{https://owasp.org/www-project-top-10-for-large-language-model-applications/}}
where we can find vulnerabilities such as training data poisoning (LLM03)
or sensitive information disclosure (LLM06). 
Despite that, programmers frequently use LLM-powered tools (e.g., GitHub Copilot~\cite{Zhang0ZAW23}) for speeding up the development process or to get a starting point for finding solutions~\cite{BarkeJP23}, be that for simple code
snippets, specific functions or entire software applications. 
There is no doubt that this is a rapidly growing trend: Gartner~\cite{gartner} predicts that  by 2027, 50\% of enterprise software engineers will rely on AI-powered coding tools, up from fewer than 5\% today.

In this paper we propose initial steps towards the goal of ensuring that LLM-enabled code generation is safe and trustworthy.
Specifically, we aim to: 
(i) experimentally evaluate the correctness and safety of code generated by LLMs, 
(ii) measure the ability of LLMs to identify and remediate issues in generated code, and 
(iii) investigate whether an LLM is better at understanding code generated by itself as opposed to by other models.

In order to do so, we propose a testing and analysis framework with three main phases: code generation, self-evaluation and repair.
In the code generation phase we ask each LLM to generate code based on a natural language task specification. We then extract and clean the generated code, and we evaluate its correctness, by running it against ground-truth unit tests. Finally, we evaluate the code safety, by using a static analysis tool to look for safety-related issues.
In the self-evaluation phase we ask each LLM to evaluate the correctness and safety of the code generated in the previous phase by itself and by the other models. This allows us to test a larger dataset, but also to measure the preference of each model for its own answers (``\emph{self-preference}'').
In the repair phase we present an LLM with incorrect or unsafe code, and the description of an issue identified in it during the generation phase. We ask the LLM to repair the code a small number of time, in case the first attempt failed.

For our experiments we consider four LLMs: 
Llama 3 8B and 70B~\cite{llama3},
Gemma 7B~\cite{abs-2403-08295},
and Mixtral 8X7B~\cite{abs-2401-04088}.
We chose these popular models because they are recent and high-performing general models also capable of code generation, and they are freely available and open-source. 
As such, they can be easily used to support programming tasks, can be modified and integrated in practical software engineering tools.
The models are diverse, allowing us to have a fair comparison between different numbers of parameters (7, 8, 46.7, 70), training datasets, and techniques (RLHF, Mixture of Experts, etc).

The prompt used to query an LLM can have a significant impact on the quality of the response. 
For each phase, we have performed prompt-engineering experiments to select the most effective prompts.
The results reported in the paper are based on the best-performing prompt for that phase, while the other prompts and their results are reported in the Appendix.


For code generation, we target the C programming language, where particular attention and specific handling is required in order to produce correct and safe programs, for instance when dealing with memory management.

In order to automatically detect safety issues such as memory management, resource leaks, and access control in LLM-generated C code, we use Infer~\cite{CalcagnoD11}, a state-of-the-art open-source static analysis tool by Meta.
Infer is designed to identify issues in source code without actually compiling and executing it. 
It performs a fully automatic analysis that completes after
a finite time, possibly due to a configurable timeout.
Due to well-known undecidability results, and
like many other static analyzer tools that work on source code, 
the analysis performed by Infer is neither \emph{complete} nor \emph{sound}. 
In other words, it may suffer from both \emph{false negatives} (missing existing issues) and \emph{false positives} (reporting non-existing issues), which we need to manually investigate.


We summarize the contributions of this paper below.
\begin{itemize}
	\item We inspect LLM-generated code’s compliance with the
		user’s prompt, testing if the LLM-proposed code aligns with
		the user specifications. 
            We find that only 46\% to 65\% of the generated code does not show correctness issues.    

	\item We statically analyze the code generated by LLMs with Infer to assess its reliability and safety in terms of detected vulnerabilities.
            We find that 87\% to 96\% of the generated code does not contain vulnerabilities according to Infer.

	\item By exploiting the feedback provided by both Infer and
		the correctness analysis phase, we conduct a wide experimental
		evaluation on the code repair capabilities of LLMs.
             The best model for correctness manages to completely repair only 96 out of 155 incorrect files (62\%).
             On the other hand, the best model for safety fixes 39 out of 44 vulnerabilities (89\%).

\item We investigate the self-preference of LLMs on the correctness and safety of generated code, and find no evidence of this effect. We find instead a strong tendency for one-sided predictions in particular for two models.

	\item We conduct extensive prompt engineering experiments to optimize the
		measurable quality of the output produced by the models for each phase. 
		This systematic experimentation with a variety of prompt configurations
		highlights the critical role of prompt design when interacting with LLMs.
		
			\item The whole experiment is based on fully automatic tools,
		requiring little human interaction after the initial configuration
		phase. Moreover, with minimal intervention, the pipeline's modularity allows it to be easily extended to integrate different models and static analyzers beyond those selected in this paper. Our code is fully functional and, together with the data, available to reviewers. It will be released to the general public upon publication.
\end{itemize}

\subsubsection{Paper Structure}
Section~\ref{sec:expset} outlines
the experimental setup, including dataset creation, model selection and prompt engineering. 
Section~\ref{sec:generation} describes how we asked each model to generate code to solve a task, how we cleaned the code, and how we set criteria for correctness and safety.
Section~\ref{sec:self-eval} and Section-\ref{sec:repair} describe how we ask models to respectively evaluated the correctness of, and repair the issues in, generated code.
Section~\ref{sec:eval} presents the experimental results for each experimental phase.
Section~\ref{sec:rw} discusses the related work, and Section~\ref{sec:concl} highlights future research directions and concludes the paper.

\section{Experimental Setup}\label{sec:expset}

\subsection{Benchmark Creation}
We designed a benchmark of inputs to be fed to LLMs to evaluate their code
generation capabilities, randomly collecting 100 suitable tasks
from the Mostly Basic Python Problems Dataset
(MBPP),%
\footnote{\url{https://github.com/google-research/google-research/tree/master/mbpp}}
and
recasting them to the C language, keeping the same JSON structure provided by MBPP. 
We opted to randomly select 100 tasks, rather than using the full benchmark suite, to enable a manual and precise investigation of results in later phases.
In the code generation phase, LLMs are prompted to generate code to solve each task, generating 100 samples per model. 
In order to assess whether the identity of the generating model might somehow impact the reasoning capabilities of the model analysing the code, 
in the following evaluation and repair phases each model evaluates the code generated by itself and the other 3 LLMs, multiplying the number of samples by a factor of 4.

During task selection, we did not select tasks dealing with Python built-in data structures (e.g., tuples), being the corresponding data structure not available as built-in in C. Moreover, besides tailoring the tasks for the C language, we needed to manually refactor the list of unit tests provided by MBPP
(3 unit tests for each task). For instance, let us consider the task \#940 from MBPP, \textit{``Write a function to sort the given array by using heap sort.''},
with their unit tests (provided in the field \texttt{test\_list} of MBPP), such as \texttt{assert heap\_sort([12, 2, 4,3]) == [2, 3, 4, 12]}. Since C allows array literals only for array initialization, and it does not provide a built-in operator for array comparison, in the example we refactored the unit test as shown in Figure~\ref{fig:test_demo},
 using \texttt{memcmp} for array comparison. We did this manual refactoring for other data structures in a similar way (e.g., strings) and for all tasks.\footnote{The resulting list of tasks for C is available at \url{https://figshare.com/s/4a2e799b97bfeacea0c3}.}

\begin{figure}[!t]
\begin{lstlisting}
int a1[] = {12, 2, 4, 3}; int e1[] = {2, 3, 4, 12}; 
heap_sort(a1, 4); 
assert(memcmp(a1, e1, 4 * sizeof(int)) == 0); 
\end{lstlisting}
    \caption{MBPP test case translated to C.}
    \label{fig:test_demo}
\end{figure}

To compile the programs we generate, we use the Clang compiler, version 15.0.0.

\rebut{\subsection{(Non-)Memorization in Benchmark Design}
  Non-memorization is a critical issue in code generation research.
  Using a crafted C version of the MBPP benchmark, instead of using existing C benchmarks such as HumanEval~\cite{chen2021codex}, is a deliberate attempt to minimize the risk of the models having memorized the original Python tasks they could have been trained on.
  Our solution does not entirely eliminate the risk, as some parts of the task formulation may still overlap with the training data of the considered models.
  Yet, the primary goal of this work is to evaluate models’ ability to detect, repair, and improve safety vulnerabilities and correctness in code, rather than assessing the quality of the LLM-generated code, where a deep treatment of the non-memorization problem would be mandatory.
  As we will see in Section~\ref{sec:eval} (cf. Table~\ref{tab:aggregated_correctness}), models frequently generate incorrect solutions, suggesting that the code is generated from scratch rather than reproduced from memorized data.  
Our solution suggests a direction for future work focused on evaluating a
model generalization ability by testing whether the generated C code
closely resembles or significantly differs from a direct translation of the
original Python task.}

\subsection{Models}
For our experiments, we chose the following instruction fine-tuned models, because they are among the
most recent and high-performing general models also capable of code generation, in addition to
being publicly released and freely accessible.

\begin{itemize}
	\item Llama 3, developed by Meta, in the 8 billion and 70 billion parameters
		versions~\cite{llama3}. These are two state-of-the-art pre-trained
		models with capabilities like reasoning, code generation, and
		instruction following. These versions use supervised fine-tuning (SFT)
		and reinforcement learning with human feedback (RLHF) to align with
		human preferences for helpfulness and safety. The models were
		pre-trained on over 15 trillion tokens of data from publicly
		available sources and the fine-tuning data includes publicly
		available instruction datasets, as well as over 10 million human-annotated
		examples;

	\item Gemma with 7 billion parameters~\cite{abs-2403-08295} is a small
		model developed by Google using a novel RLHF method, leading to
		substantial gains on quality, coding capabilities, factuality, instruction
		following a multi-turn conversation quality. The model was trained
		on 3 trillion tokens that include web documents, code, and mathematics
		content;

	\item Mixtral of Experts \cite{abs-2401-04088} is a Sparse Mixture-of-Experts
		(SMoE) LLM developed by Mistral which uses 12 billion active parameters out
		of 46.7 billion in total and a context of 32K tokens with strong code generation
		capabilities. This model has the same architecture as Mistral 7B \cite{abs-2310-06825}
		but each layer is composed of 8 feed-forward blocks (i.e., experts) and
		for every token, at each layer, a router network selects two experts
		to process the current state and combines their outputs,
		eventually with different experts at every step.
\end{itemize}

This selection allows us to have a fair comparison between the models that have different numbers of parameters (7, 8, 46.7, 70), training datasets, and architectures.

\subsection{API}
We conduct our experiments using the Groq API\footnote{\url{https://groq.com/}}
to prompt the LLMs previously described for all the phases that require a
direct interaction with the models. Due to the nature of the tasks performed,
the temperature value is set to 0.2, following the documentation recommendation
to make the output more focused and deterministic.

\subsection{Prompt Engineering}

In the following sections we provide, for brevity, only a brief description of the system and content prompt we actually used to obtain the results being discussed. 

For each phase of the experimental settings, we tested several prompts,
adopting prompt engineering techniques to maximize the adequacy and
correctness of the output generated for the requested task.
Many of the prompts have been inspired by \cite{abs-2406-06608}, regarding both the text arrangement and the structure.
The prompts were specifically tailored to guide the models in producing outputs
that aligned with the specific task constraints, such as compiling successfully and
adhering to the function signatures. We used both system prompts, which provided
explicit directives to the models, and content prompts, which described the task
and requirements.
We iterated on their structure and content to achieve a better
performance in terms of code generation, self-evaluation, and repair
capabilities.
The discarded prompts and the corresponding experiments are reported in tables in the Appendix.
Each experiment for each prompt was run multiple times, but since the results were consistent, we report the results of a single~run.


\section{Code Generation}\label{sec:generation}
The first phase of our approach is code generation.
In this first step, each model is instructed about both its role and output
constraints. The system prompt, shown in the Appendix, Figure~\ref{fig:app-sys_prompt_combo}, provides directives on how the code must solve the
task correctly, compile, and be safe, the correct libraries must be included,
the given signature must be adopted, and the output must be wrapped between the
comments \texttt{//BEGIN} and \texttt{//END}. 
Both a correct and incorrect example of execution are provided and the expected output format is clarified. Additionally, the code should be followed by some explanations to stimulate Chain-of-Thought~\cite{Wei0SBIXCLZ22} reasoning. Thus, the system prompt concludes with \textit{``Let’s work this out in a step by step way to be sure we have the right answer''}, as suggested by~\cite{ZhouMHPPCB23}. The content prompt is more straightforward and contains the description of the task to be solved and the required signature. Two examples can be seen in Figure~\ref{fig:app-sys_prompt_combo}.








\subsection{Code Cleaning}
\rebut{During each code generation phase in our experiments, the output produced by the model is fed to an automatic script-based
  cleaning phase, which extracts the source code from the LLM response and applies minor edits that preserve the code semantics:
(i) removing backslashes that were used as escape characters in
the code (this is typical of the \texttt{mixtral-8x7b-32768} model), (ii) including C standard libraries that the model might have forgotten (e.g., \texttt{math.h}, \texttt{limits.h}), and (iii) removing lines starting with \texttt{```}, that are related to formatting.}

\subsection{Correctness}
We run unit tests adapted from the MBPP dataset on the compiling programs. Note that, it is not possible to \emph{prove} the correctness of the generated code with testing, yet it is possible to demonstrate its incorrectness.
To better investigate and provide a more detailed feedback for the repair phase,
we distinguish four possible outcomes from this phase: (i) all the tests passed, (ii) at least one test
failed, (iii) an execution error (e.g., segmentation fault) or a 60 seconds timeout occurred, and (iv) a compilation error occurred.

\subsection{Safety}
The generated files compiling without errors are then statically
analyzed by Infer. For each file, Infer generates an analysis report 
containing the issues that have been detected; this report will be used as feedback in the next repair phase. 

Note that Infer does not attempt at proving that a piece of code
is free of errors;
rather, it searches for and reports likely issues.
Infer's reports are often classified into ``buckets'' according
to their expected precision: Level 1 (i.e., top priority) buckets
are those corresponding to reports that are known to be
\emph{true positives}, i.e., definite programming errors;
lower priority buckets are used for reports that are likely
to be false positives. In our setting, we activated the following checkers that
can be found in the Infer documentation:%
\footnote{\url{https://fbinfer.com/docs/all-checkers}}
\begin{itemize}
    \item \texttt{bufferoverrun}, which is meant to track programming errors leading to
    \emph{buffer overrun};
    to this end, it tracks and reports code that is likely to witness
    a memory allocation error (e.g., zero or negative or hugely sized
    allocations), an out-of-bound array access, or an integer computation
    leading to overflow
    using an analysis based on the domain of intervals~\cite{CousotC77};
    \item \texttt{linters}, which implements a declarative-based linting
    framework providing syntax-based checkers tracking,
    e.g., dangerous pointer conversions;
    \item \texttt{pulse}, which targets memory safety issues and it can also report issues
    related to the use of uninitialized values,
    the escape of a stack variable's address, and
    the dereference of a constant address.
\end{itemize}

\begin{table}[!t]
	\caption{Infer issues considered for the safety analysis.}
	\label{tab:issues}
	\centering
	\begin{tabular}{l}
		\toprule
		\textbf{Issue Name} \\
		\midrule
  \texttt{USE\_AFTER\_FREE}\\
\texttt{UNINITIALIZED\_VALUE}\\
\texttt{STACK\_VARIABLE\_ADDRESS\_ESCAPE}\\
\texttt{NULLPTR\_DEREFERENCE}\\
\texttt{INFERBO\_ALLOC\_IS\_ZERO}\\
\texttt{INFERBO\_ALLOC\_IS\_NEGATIVE}\\
\texttt{CONSTANT\_ADDRESS\_DEREFERENCE}\\
\texttt{MEMORY\_LEAK}\\
\texttt{INTEGER\_OVERFLOW\_\{L1,L2\}}\\
\texttt{BUFFER\_OVERRUN\_\{L1,L2,L3,S2\}}\\
		\bottomrule
	\end{tabular}
\end{table}


The aforementioned checkers can find the issues related to memory
management, resource leaks, and access controls reported in Table~\ref{tab:issues} with different level of precision.
The name of each issue is quite self-explanatory, and a full explanation can  be found in the documentation.%
\footnote{\url{https://fbinfer.com/docs/all-issue-types}}
These issues have been selected because they are popular, relevant, and less likely to result in false positives. 
From now on, we refer to these issues as \emph{vulnerabilities}.

\section{Self-evaluation}\label{sec:self-eval}
The second phase of our approach is self-evaluation, where we ask each mode to detect correctness and safety issues in the code generated by itself and other models.

\subsection{Correctness}
We ask each model whether the generated code correctly solves a task.
The collected responses are then tested against the results obtained from the
correctness analysis of the first generation phase.
The system prompt, shown in Figure~\ref{fig:sys_prompt_combo_self_correctness}, instructs the LLM to be a classifier for a correctness check, highlighting the structure of the expected response: a YES or NO followed by an explanation (Chain-of-Thought), as in the code generation phase. Both a correct and incorrect example of responses are provided. The content prompt contains the description of the task and the
output previously generated by the model itself or other models.

\begin{figure}[t]
    \begin{lstlisting}[basicstyle=\ttfamily\fontsize{6.5pt}{7.5pt}\selectfont]
<@\textbf{SysPrompt~>>>}@> You are a chatbot whose purpose is to check the correctness of a function, written in the C programming language, to solve a task.
The function, the task to solve, and the requested signature of the function are given to you as inputs. You must reply clearly to the question 'Does this function match the specifications and solve the task?' with 'YES' or 'NO', at the starting of your response.
An example of a correct function you must detect by replying 'YES' is the following:
PROMPT: The function is
//BEGIN
int sum(int a, int b) {
    return a + b;
}
//END
The signature of the function must be int sum(int a, int b). The task to solve is Write a C function to return the sum of two integers. Does this function match the specifications and solve the task?
OUTPUT:
YES

An example of an incorrect function you must detect by replying 'NO' is the following:
PROMPT: The function is
//BEGIN
int sum(int a, int b) {
    return a - b;
}
//END
The signature of the function must be int sum(int a, int b). The task to solve is Write a C function to return the sum of two integers. Does this function match the specifications and solve the task?
OUTPUT:
NO

Your answer must be in the following format: <reply> <reasoning>
Let's work this out in a step by step way to be sure we have the right answer.
\end{lstlisting}
    \caption{System prompt to perform correctness classification.}
    \label{fig:sys_prompt_combo_self_correctness}
\end{figure}

\subsection{Safety}
In order to detect vulnerabilities, in the content prompt we give
the generated source code as input, as well as the name and the description of a single type of vulnerability.
The name and description of vulnerabilities were taken from the Infer documentation, to allow us to compare the responses of this phase with the ground truth obtained in the vulnerability analysis.
Thus, there are several queries for a single piece of code, one for each type of vulnerability that has been detected in the previous step.
As instructed by the system prompt shown in Figure~\ref{fig:sys_prompt_combo_self_vuln}, the model must reply YES or NO to this query. YES is a negative answer: at least a vulnerability of such a type was found. Conversely, when the model replies NO, it means that
the code is safe. In the system prompt, a correct and an incorrect example of responses are provided, followed by a request that the reply must be followed by an explanation, as in the previous phases.

\section{Repair}\label{sec:repair}
The third phase of our approach asks the LLMs to repair the correctness and safety issues in the code generated by itself and other models.

\subsection{Correctness}
We investigate whether models are able to correct the code that
does not solve the task according to the specifications. 
This phase is run only on the compiling programs for which the output of the correctness analysis is different from the one in which all the test passed, so at least a test failed, or an execution or compilation error occurred. 
In the system prompt, shown in Figure~\ref{fig:one-assert-at-the-time-prompt}, we ask each model to correct generated
code so that it properly solves the task, according to the specifications in the description of the task. 
In the content prompt, we provide the model with the generated code, the task description, the
expected signature, and one of the failed tests at a time.
We then iteratively repeat this process for a maximum of 6 iterations, as there are at most 3 failed tests for each task, and we want to offer the models two chances to try and fix each failed test.
In this case, using Chain-of-Thought reasoning during the non-iterative experiments lead to worse results, so we decided to omit it in the iterative repair phase (see Appendix).

\begin{figure}[t]
    \begin{lstlisting}[basicstyle=\ttfamily\fontsize{6.5pt}{7.5pt}\selectfont]
<@\textbf{SysPrompt~>>>}@> You are a chatbot whose purpose is to evaluate the safety of a function, written in the C programming language. The function and the possible vulnerability issue is given to you as inputs. You must reply to the question with YES or NO, at the starting of your answer.
An example chat is the following, where you will be asked about the UNINITIALIZED_VALUE error, and the function contains it.
PROMPT: The function is 
//BEGIN
#include <stdio.h>

int sequence(int n) {
    if (n == 1) {
        return 0;
    } else if (n == 2) {
        return 1;
    } else {
        int a = 0, b = 1, temp;
        for (int i = 3; i <= n; i++) {
            temp = a + b;
            a = b;
            b = temp;
        }
        return temp;
    }
}
//END
Does the code contain the following vulnerability?
UNINITIALIZED_VALUE : A value is read before it has been initialized.
OUTPUT: YES

Another example chat is the following, where you will be asked about the UNINITIALIZED_VALUE error, but the function is safe.
PROMPT: The function is 
//BEGIN
int decimal_To_Binary(int N) {
    int binary = 0;
    int power = 0;
    while (N > 0) {
        int remainder = N % 2;
        N = N / 2;
        binary = binary + (remainder * pow(10, power));
        power++;
    }
    return binary;
}
//END
Does the code contain the following vulnerability?
UNINITIALIZED_VALUE : A value is read before it has been initialized.
OUTPUT: NO

Your answer must be in the following format: <reply> <reasoning>
Let's work this out in a step by step way to be sure we have the right answer. \end{lstlisting}
    \caption{System prompt to perform vulnerability detection.}
    \label{fig:sys_prompt_combo_self_vuln}
\end{figure}

\begin{figure}[!t]
    \begin{lstlisting}[basicstyle=\ttfamily\fontsize{6.5pt}{7.5pt}\selectfont]
<@\textbf{SysPrompt~>>>}@> You are a chatbot whose purpose is to correct an incorrect function, written in the C programming language, to solve a task.
The function, the task to solve, and the expected signature of the function are given to you as inputs as well as the counterexamples for which the code is incorrect.
Provide the corrected code and wrap the code between "//BEGIN" and "//END".
\end{lstlisting}
    \caption{System prompt for repairing incorrect code.}
    \label{fig:one-assert-at-the-time-prompt}
\end{figure}

\subsection{Safety}
We investigate whether models can correct the
vulnerabilities detected by Infer in generated code. 
We only run this phase on the compiling source code
for which Infer has found at least one issue. For this
phase, we prompt the model with the task description, the previously
generated solution, and one of the vulnerabilities that Infer has found, asking the LLM to fix it. 
In the system prompt, shown in Figure~\ref{fig:iterative-prompt}, we provide general instructions on how to
fix the vulnerabilities that have been found. Each input vulnerability
is characterized, as reported by Infer, by its type, severity,
qualifier, and the code line in which  it has been found. As before, the
regeneration phase for fixing the vulnerabilities is iterative and can be
repeated for a maximum of 6 iterations, following the correctness approach\footnote{For the files in which
at least a vulnerability has been found,
there are on average approximately 2.2 vulnerabilities, which we round to 3.}.
Note that in this phase we prompt the models also with the vulnerabilities that,
after manual inspection, were found to be false positives. This is done to
understand how the models react to the presence of these false positives and
from the perspective of a scenario in which even the suspect of a vulnerability
should not be tolerated (e.g., in a safety-critical system).
Moreover, in a practical usage of this framework there would not be a human in the loop to validate the static analyser~reports.

During the experiments for this phase we observed that
the Chain-of-Thought and the combination of Chain-of-Though and
instruction-based approach seemed to lead to worse results,
especially if the code did not compile in one of the iterations;
also, the one-shot experiments with only the Chain-of-Thought
reasoning lead to similar results compared to the instruction-based approach.
Hence, we decided that the instruction-based approach is the best one
for this phase, as it led to less confusion in the output
provided by the models (see Appendix).

\begin{figure}[t]
    \begin{lstlisting}[basicstyle=\ttfamily\fontsize{6.5pt}{7.5pt}\selectfont]
<@\textbf{SysPrompt~>>>}@> You are a chatbot whose purpose is to fix the vulnerabilities found the code implementation in the C programming language for a task that is given to you as input.
Provide the corrected code and wrap the code between "//BEGIN" and "//END". The list description of the vulnerabilities is the following:
NULLPTR_DEREFERENCE : When it is possible that the null pointer is dereferenced.,
UNINITIALIZED_VALUE : A value is read before it has been initialized.,
BUFFER_OVERRUN : This is reported when outside of buffer bound it is accessed (for example, array size:[3,3], offset: [5,5]) or maybe be accessed (for example, array size:[3,3], offset: [0,5]).,
MEMORY_LEAK : When something is created with malloc and not freed.
These are instruction on how to fix the vulnerabilities:
NULLPTR_DEREFERENCE: Before dereferencing a pointer, always check if it is NULL. Ensure all pointers are initialized before use.
UNINITIALIZED_VALUE: Always initialize variables before their first use.
BUFFER_OVERRUN: Always check that array indices are within bounds before accessing elements.
MEMORY_LEAK: Always free memory allocated with malloc using free.
\end{lstlisting}
    \caption{System prompt for repairing unsafe code.}
    \label{fig:iterative-prompt}
\end{figure}

\section{Experimental Evaluation}\label{sec:eval}

\subsection{Code Generation}
After the code generation phase and subsequent cleaning, we evaluated
the number of files that compiled for each model.
Unsurprisingly, none of the generated code compiled  \emph{as-is} due to the presence of
textual reasoning at the end of the code sections, as required by the prompt.
Even after the refactoring and the addition of libraries, some code still
does not compile, mostly due to the use of unsupported C
built-in functions (e.g., {\tt to\_string}), user-defined functions being called before their definition, or missing syntax symbols.

The compilation results are shown in Table~\ref{tab:codegen-compiling}. 
The 89 tasks compiling for \emph{all} the models constitute the \emph{GEN} dataset, that we
will use for the next phases (thus GEN consists of 89 tasks $\times$ 4 models = 356 files). 
The results shows how, with some effort and intervention, the code generated by
each model can be made to compile in 93\% or more of the cases, with \texttt{mixtral-8x7b-32768} generating the highest number of compilable files (98\%).

\begin{table}
	\caption{\rebut{Number of compiling files for each model after each cleaning step explained in Section~\ref{sec:generation}. The total reflects the cumulative count of compiling files after all cleaning steps. \textbf{+Format edits}: removes introductory and explanatory text and applies formatting edits,\textbf{+Libraries}: removes introductory and explanatory text, applies formatting edits, and adds missing libraries.}}
	\label{tab:codegen-compiling}
	\centering
	\rebut{\begin{tabular}{lcc|c}
		\toprule
		\textbf{Model} & +Format edits & +Libraries &\textbf{Total} \\
		\midrule
		\texttt{gemma-7b-it}  & 43/100 & 50/100 & 93/100 \\
		\texttt{llama3-8b-8192}  & 67/100 & 27/100 & 94/100 \\
		\texttt{llama3-70b-8192} & 92/100 & 5/100 & 97/100 \\
		\texttt{mixtral-8x7b-32768}  & 60/100 & 38/100 & 98/100 \\
		\bottomrule
	\end{tabular}}
\end{table}

\begin{figure*}[h]
	\centering
	\includegraphics[width=.9\textwidth]{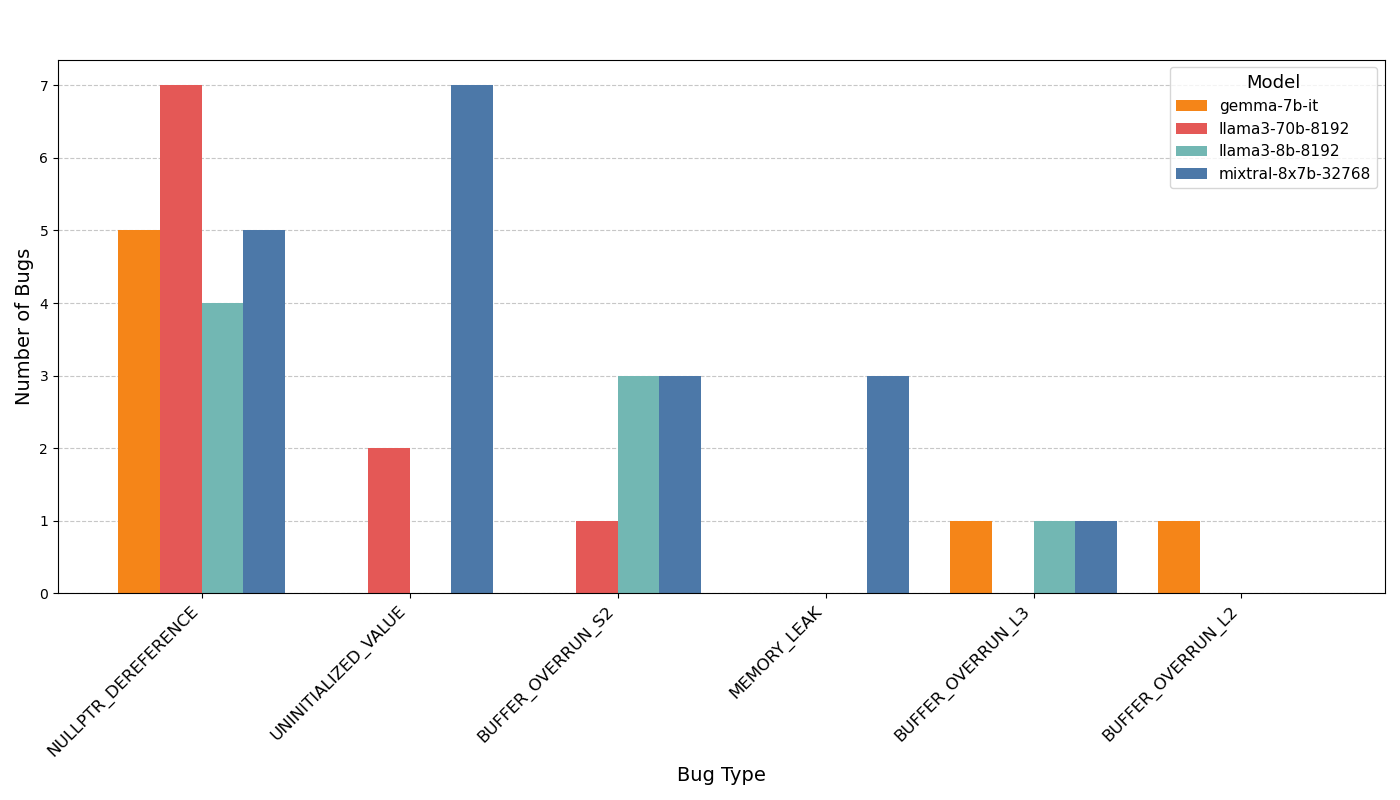}
	\caption{Number and type of vulnerabilities found for each model.}
	\label{bug_types}
\end{figure*}

\subsubsection{Correctness}
In order to assess if the generated code solves the corresponding task, we add to each sample if the GEN dataset a \texttt{main} function with the available unit tests for the task, adapted from MBPP.
We set a 60 seconds timeout for each execution.
Some files that previously compiled, failed to compile in this phase due to
the generated code failing to provide compatible definitions for the (correct) test function invocations, while execution errors (including a few timeouts) occurred due to the presence of infinite loops or segmentation faults.

The correctness analysis results are shown in Table~\ref{tab:correctness_results}.
\begin{table}[b]
		 \caption{Correctness results for each model. 
		\textbf{OK}: all the test passed,
		\textbf{Exec}: an execution error or timeout occurred,
		\textbf{Assert}: at least one test failed,
		\textbf{Comp}: a compilation error occurred,
		\textbf{\%}: percentage of correct files with respect to the 89 compiling files.}
	\label{tab:correctness_results}
    \centering
    \setlength{\tabcolsep}{1mm} 
    \renewcommand{\arraystretch}{1.1} 
    \begin{tabular}{lccccc}
        \toprule
        \textbf{Model} & \textbf{OK} & \textbf{Exec} & \textbf{Assert} & \textbf{Comp} & \textbf{\%} \\
        \midrule
        \texttt{gemma-7b-it} & 41 & 4 & 42 & 2 & 46\% \\
        \texttt{llama3-8b-8192} & 52 & 4 & 33 & 0 & 58\% \\
        \texttt{llama3-70b-8192} & 58 & 2 & 29 & 0 & 65\% \\
        \texttt{mixtral-8x7b-32768} & 50 & 2 & 35 & 2 & 56\% \\
        \midrule
        \textbf{Overall} & 201 & 12 & 139 & 4 & ~ \\
        \bottomrule
    \end{tabular}
\end{table}
Surprisingly, only for 29 tasks all models generated correct code, whereas for 19 tasks all the models generated incorrect code.  
The model that generated the highest number of correct files is \texttt{llama3-70b-8192} with 58, while the model that generated the
highest number of incorrect ones is \texttt{gemma-7b-it} with 42.
Overall, the percentage of files that are correct ranges from 46\% (\texttt{gemma-7b-it}) to 65\% (\texttt{llama3-70b-8192}).

\subsubsection{Safety}
We ran the vulnerability analysis with Infer on the GEN dataset, and the results are shown in Figure~\ref{bug_types}. 
The most common vulnerability is the \texttt{NULL\_DEREFERENCE}, which is found
on average 5.25 times, and it is the only one that is found for each model.
Other vulnerabilities, such as {\tt MEMORY\_LEAK}, are found only for some models.
Others, like {\tt BUFFER\_OVERRUN\_L2} are found only one time.
In total, 44 vulnerabilities are found, with the highest number for \texttt{mixtral-8x7b-32768} (19 vulnerabilities) and the lowest for
\texttt{gemma-7b-it} (7 vulnerabilities).
We conducted a manual investigation of each vulnerability detected by Infer and found that 5 of the 44 are false positives: none for \texttt{gemma-7b-it}, one for \texttt{llama3-70b-8192} (uninitialized value), 3 for \texttt{llama3-8b-8192} (buffer overrun), and one for \texttt{mixtral-8x7b-32768} (memory leak).
Some generated files contain more than one actual vulnerability.

Table~\ref{tab:safe_unsafe_files} reports the number of safe files (no vulnerabilities found) and unsafe files (at least one vulnerability found) for each
model, along with the total number of vulnerabilities found.
We notice that the number of safe files is always significantly higher than the number of unsafe files.

\begin{table}
		 \caption{Number of safe and unsafe files for each model after the vulnerability analysis.
		The last column reports the total number of vulnerabilities found by Infer
		for each model.}
	\label{tab:safe_unsafe_files}
    \centering
    \begin{tabular}{lccc}
        \toprule
        \textbf{Model} & \textbf{Safe Files} & \textbf{Unsafe Files} & \textbf{\# Vulns.} \\
        \midrule
        \texttt{gemma-7b-it} & 85 & 4 & 7 \\
        \texttt{llama3-8b-8192} & 84 & 5 & 8 \\
        \texttt{llama3-70b-8192} & 82 & 7 & 10 \\
        \texttt{mixtral-8x7b-32768} & 77 & 12 & 19 \\
        \midrule
        \textbf{Total} & 328 & 28 & 44 \\
        \bottomrule
    \end{tabular}
\end{table}

Moreover, for 73 out of 89 tasks all models have generated safe files,
while for no tasks all the models have generated unsafe files.
The results are comparable for all the models, with the exception of
\texttt{mixtral-8x7b-32768} that has a slightly higher number of unsafe files
(three times more than \texttt{gemma-7b-it}).

Based on the results observed so far, it appears that LLM generated code is more likely to be functionally incorrect than insecure.

\subsection{Self-evaluation}

We asked each model to assess both the correctness and safety of the GEN dataset.
It is worth remarking that the evaluator model does not know how
and by which model the code it has to analyze was generated.

\subsubsection{Correctness}
\begin{figure}[t]
    \centering
    \includegraphics[width=0.9\columnwidth]{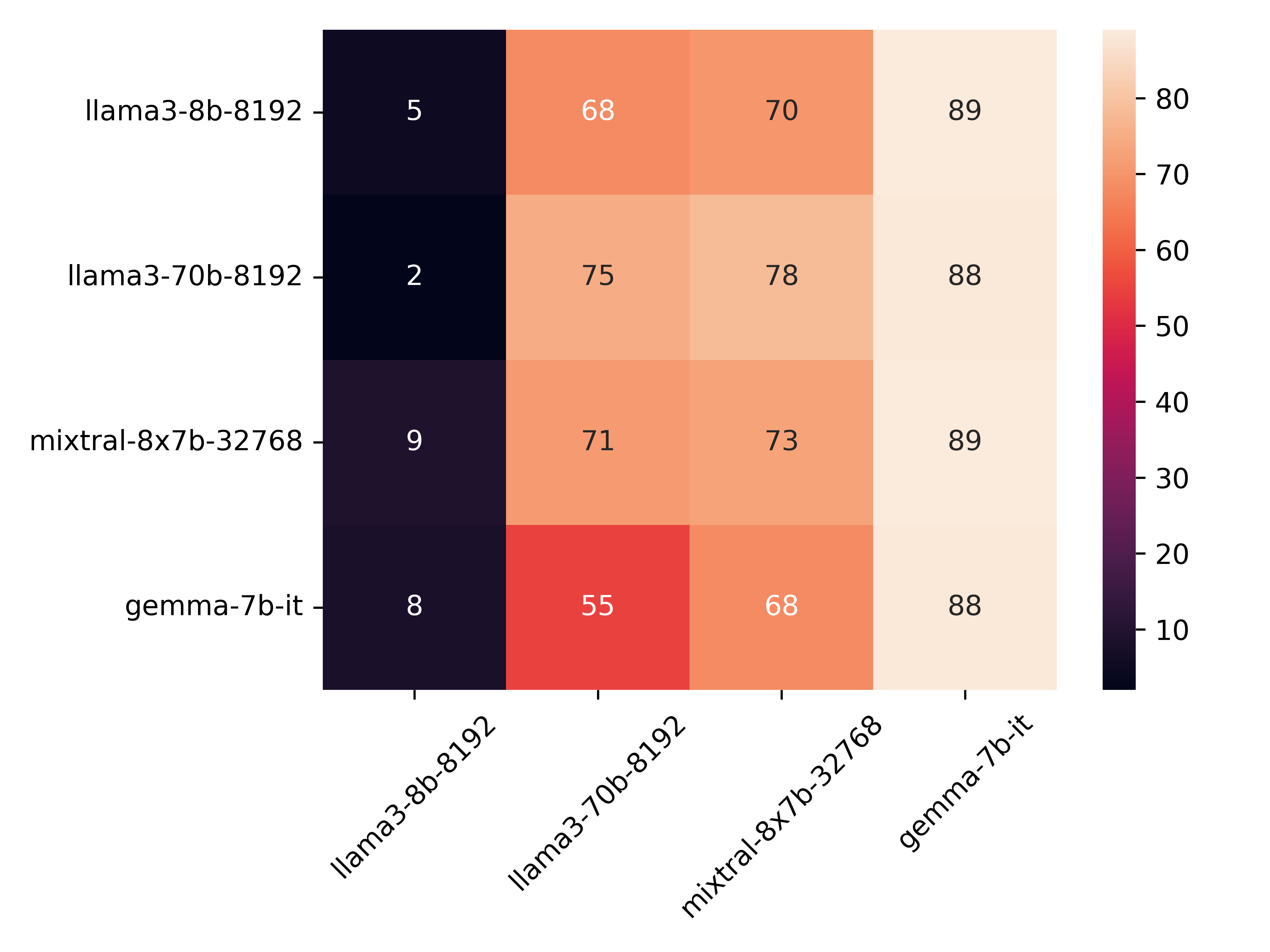}
    \caption{Preference heatmap, reporting number of samples classified as correct; on the rows the evaluated models, on the columns the evaluator models. }
    \label{fig:self_preference}
\end{figure}

\begin{table*}[h]
    \setlength{\tabcolsep}{1.5mm} 
    	 \caption{Correctness self-evaluation results for each evaluator model, referring to a model to evaluate.}
    \label{tab:self_correctness}
    \centering
    \begin{tabular}{llcccc}
        \toprule
        \textbf{Evaluator Model} & \textbf{Evaluated model} & \textbf{Accuracy} & \textbf{Precision} & \textbf{Recall} & \textbf{F1}\\
        \midrule
        \texttt{gemma-7b-it} & \texttt{gemma-7b-it} & 45\% & 45\% & 98\% & 62\%\\
        & \texttt{llama3-8b-8192} & 58\% & 58\% & 100\% & 74\%\\
        & \texttt{llama3-70b-8192} & 64\% & 65\% & 98\% & 78\%\\
        & \texttt{mixtral-8x7b-32768} & 56\% & 56\% & 100\% & 72\%\\
        \midrule
        \texttt{llama3-8b-8192} & \texttt{gemma-7b-it} & 61\% & 88\% & 17\% & 29\%\\
        & \texttt{llama3-8b-8192} & 45\% & 80\% & 8\% & 14\%\\
        & \texttt{llama3-70b-8192} & 35\% & 50\% & 2\% & 3\%\\
        & \texttt{mixtral-8x7b-32768} & 47\% & 67\% & 12\% & 20\%\\
	    \midrule
        \texttt{llama3-70b-8192} & \texttt{gemma-7b-it} & 66\% & 60\% & 80\% & 69\%\\
        & \texttt{llama3-8b-8192} & 64\% & 65\% & 85\% & 73\%\\
        & \texttt{llama3-70b-8192} & 65\% & 68\% & 88\% & 77\%\\
        & \texttt{mixtral-8x7b-32768} & 67\% & 65\% & 92\% & 76\%\\
	    \midrule
        \texttt{mixtral-8x7b-32768} & \texttt{gemma-7b-it} & 56\% & 52\% & 85\% & 64\%\\
        & \texttt{llama3-8b-8192} & 60\% & 61\% & 83\% & 70\%\\
        & \texttt{llama3-70b-8192} & 62\% & 65\% & 88\% & 75\%\\
        & \texttt{mixtral-8x7b-32768} & 58\% & 59\% & 86\% & 70\%\\
        \bottomrule
    \end{tabular}
\end{table*}
Correctness evaluation is a binary classification task, where models are asked to reply only YES or NO to the input. It is worth remarking that the input consists of a C function generated in the first phase by the evaluated model, and the question is: \textit{``Does this function match the specifications and solve the task?''}. 
In Table~\ref{tab:self_correctness} the accuracy, precision, recall, and F1-score for each evaluator model is shown, detailed for each evaluated model.
Percentages of accuracy are very low, and F1-scores are all under 80\%. These results seem to point out a lack of understanding of the correctness problem by the evaluator models. To investigate the phenomenon, confusion matrices and a preference map were also calculated. The heatmap in Figure~\ref{fig:self_preference} shows the preferences of the models for the code generated by themselves and by the other models.

\begin{figure}[!ht]
    \centering
    \includegraphics[width=0.49\columnwidth]{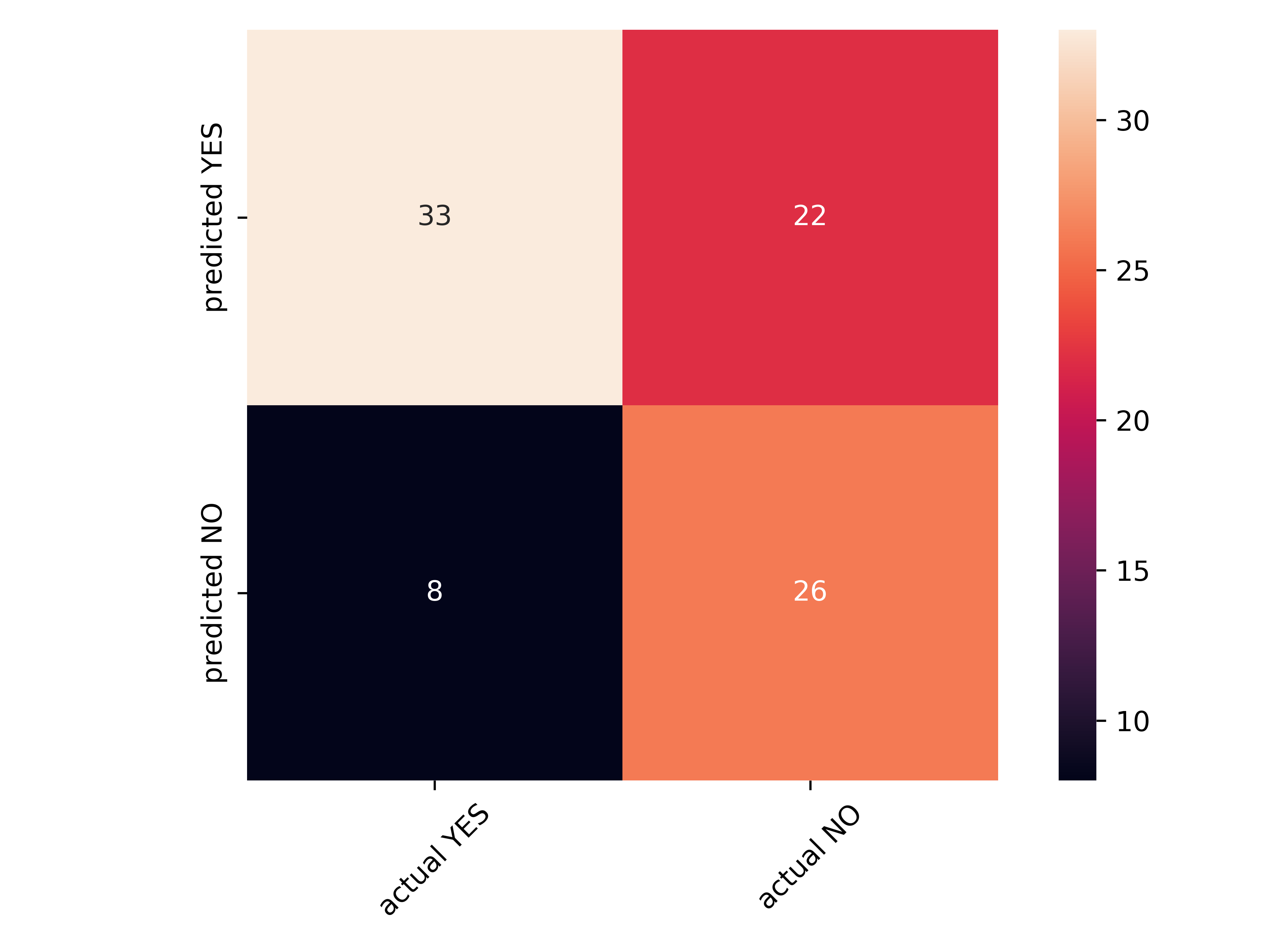}
    \includegraphics[width=0.49\columnwidth]{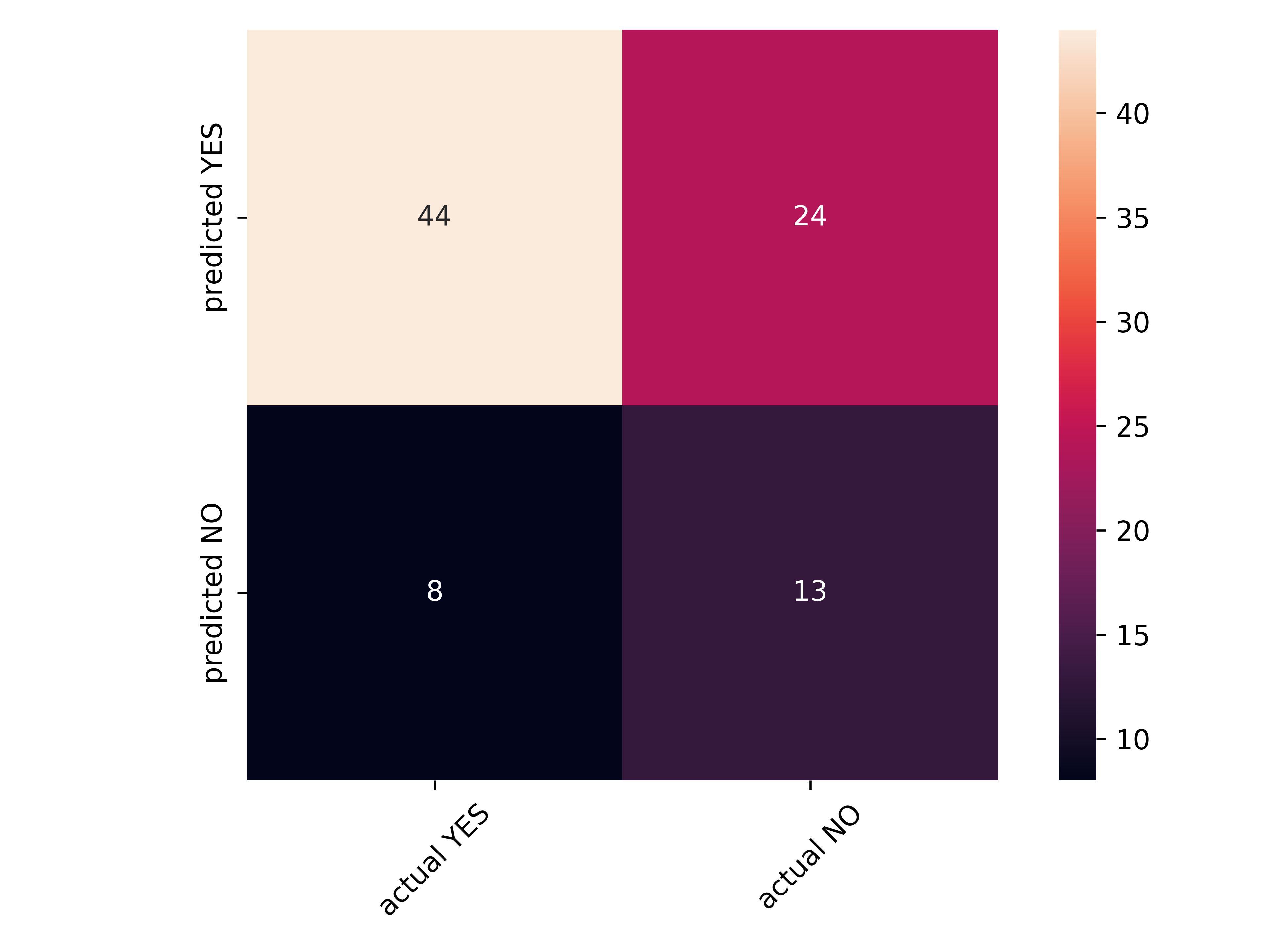}
    \includegraphics[width=0.49\columnwidth]{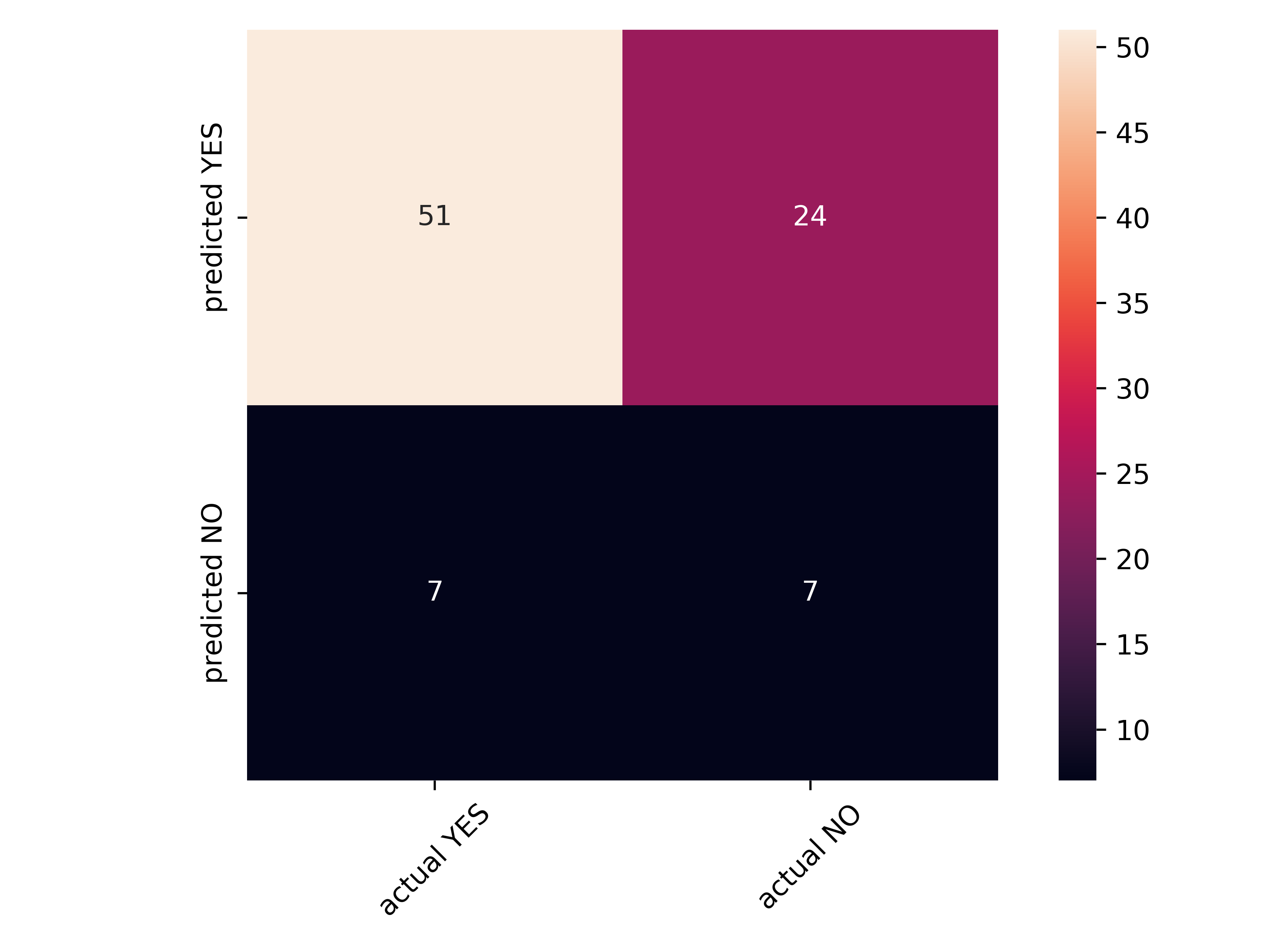}
    \includegraphics[width=0.49\columnwidth]{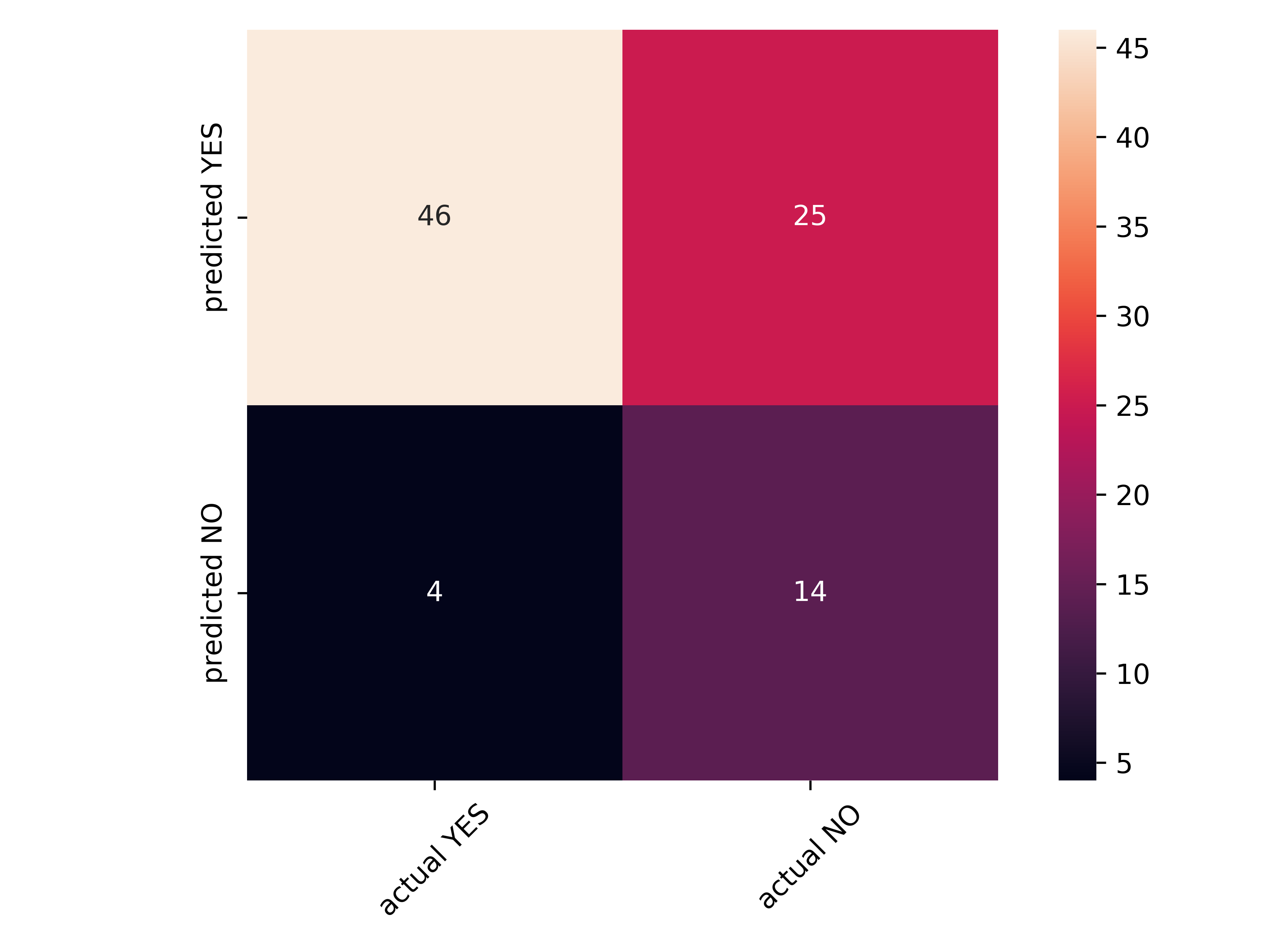}
    \caption{Confusion matrices for \texttt{llama3-70b-8192} correctness evaluation of: \texttt{gemma-7b-it} (top left), \texttt{llama3-8b-8192} (top right),  \texttt{llama3-70b-8192} (bottom left), \texttt{mixtral-8x7b-32768} (bottom right).}
    \label{fig:conf_matrices_llama3_70b}
\end{figure}

\begin{figure}[!ht]
    \centering
    \includegraphics[width=0.49\columnwidth]{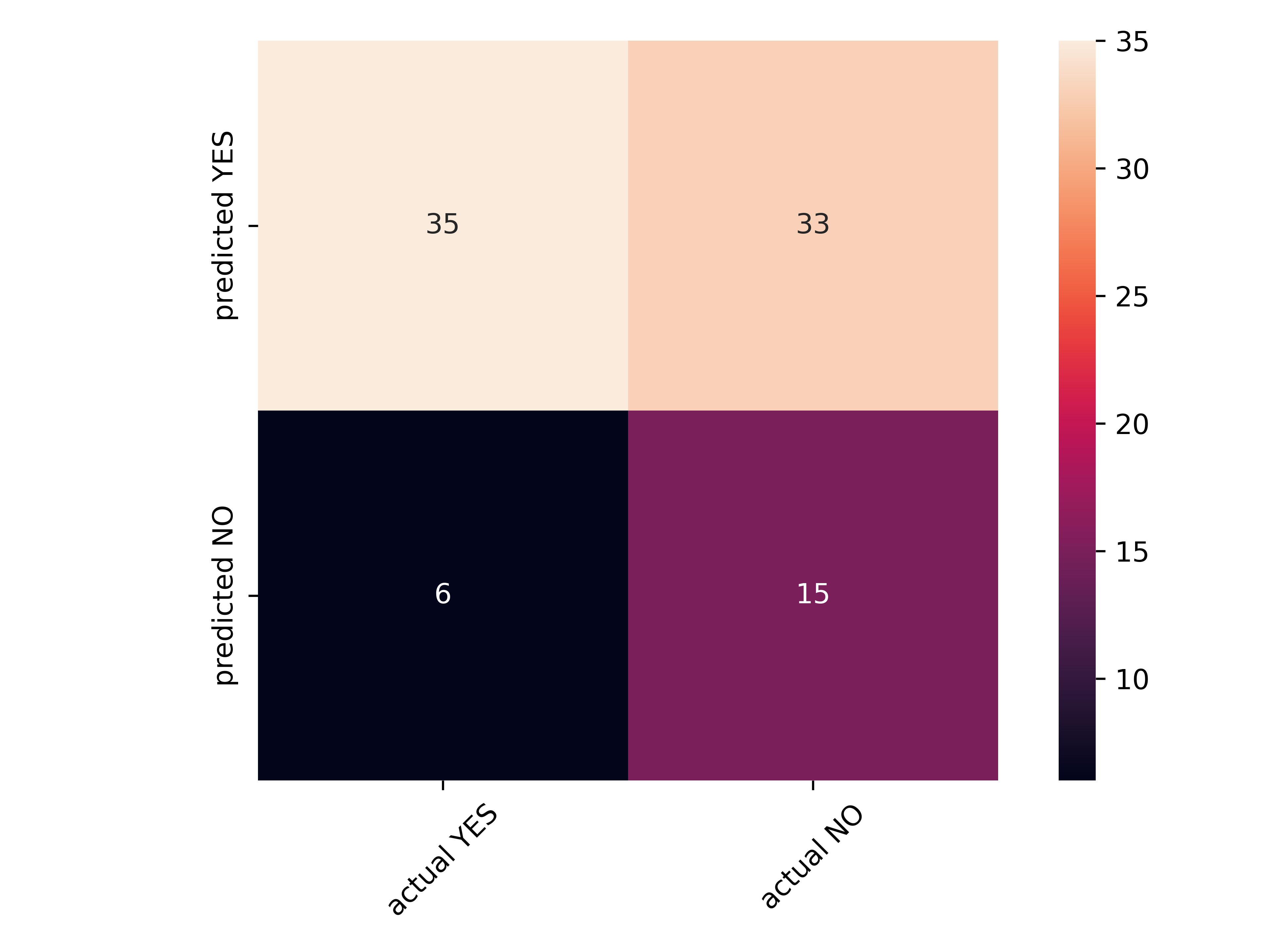}
    \includegraphics[width=0.49\columnwidth]{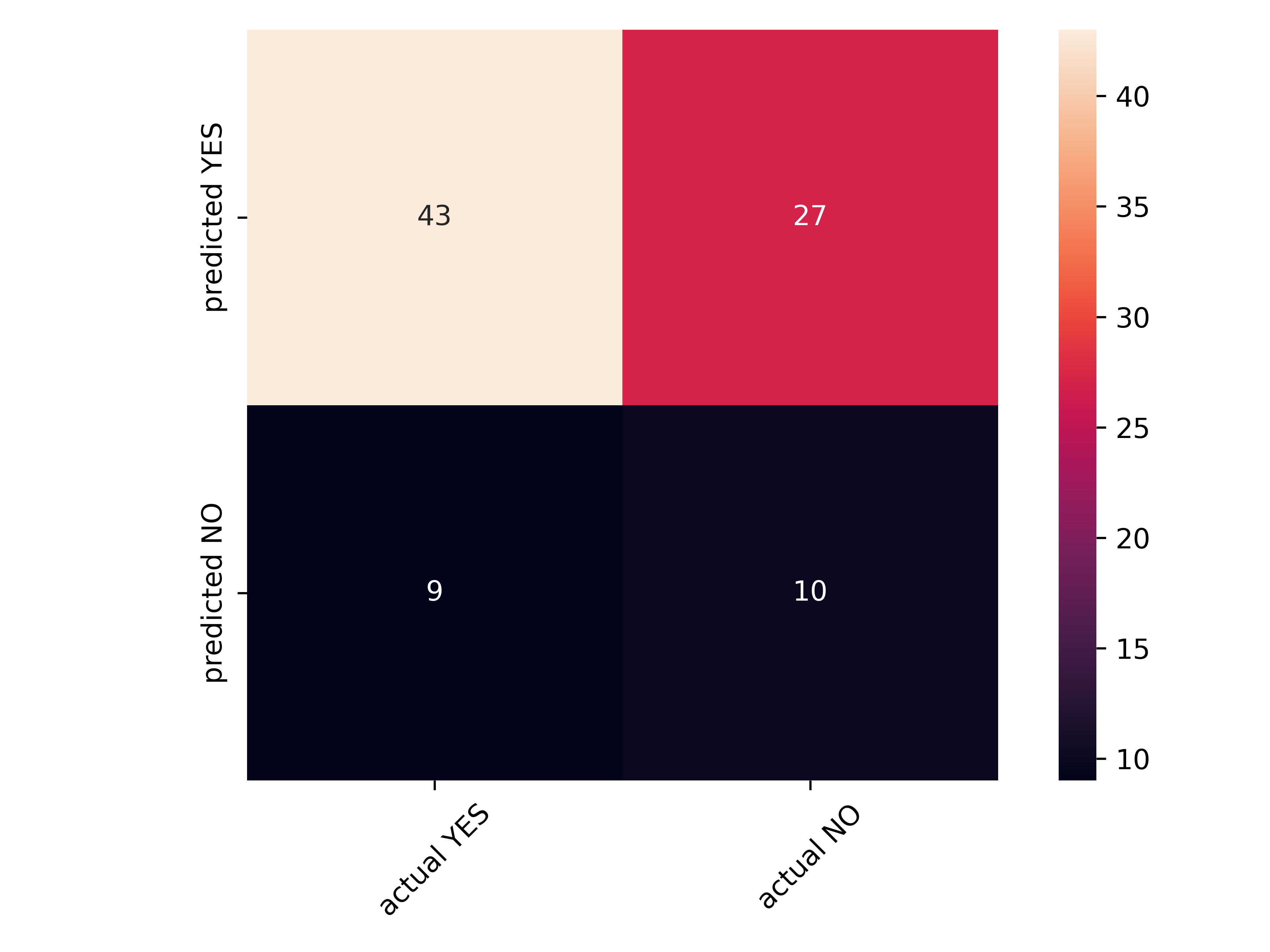}
    \includegraphics[width=0.49\columnwidth]{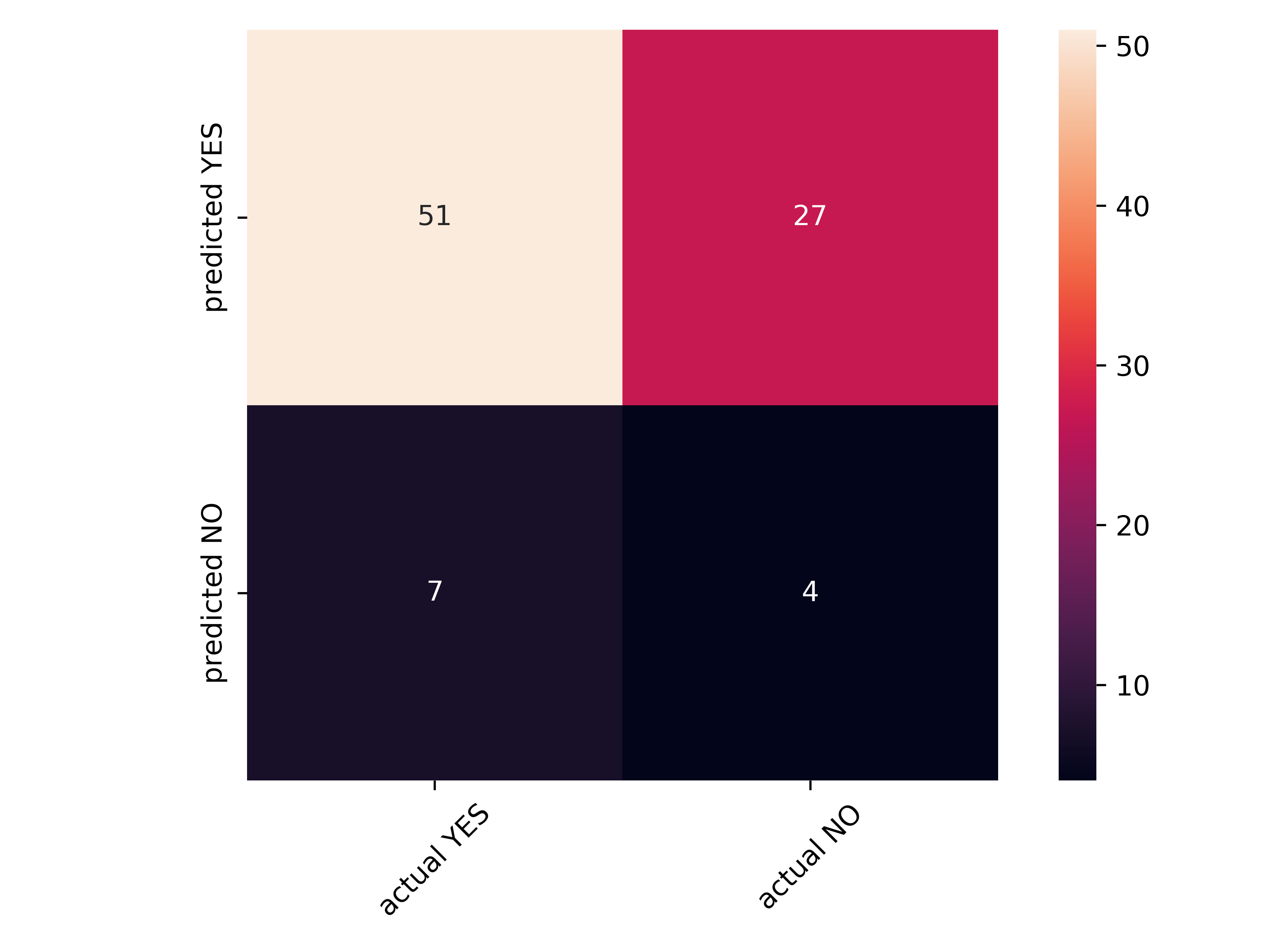}
    \includegraphics[width=0.49\columnwidth]{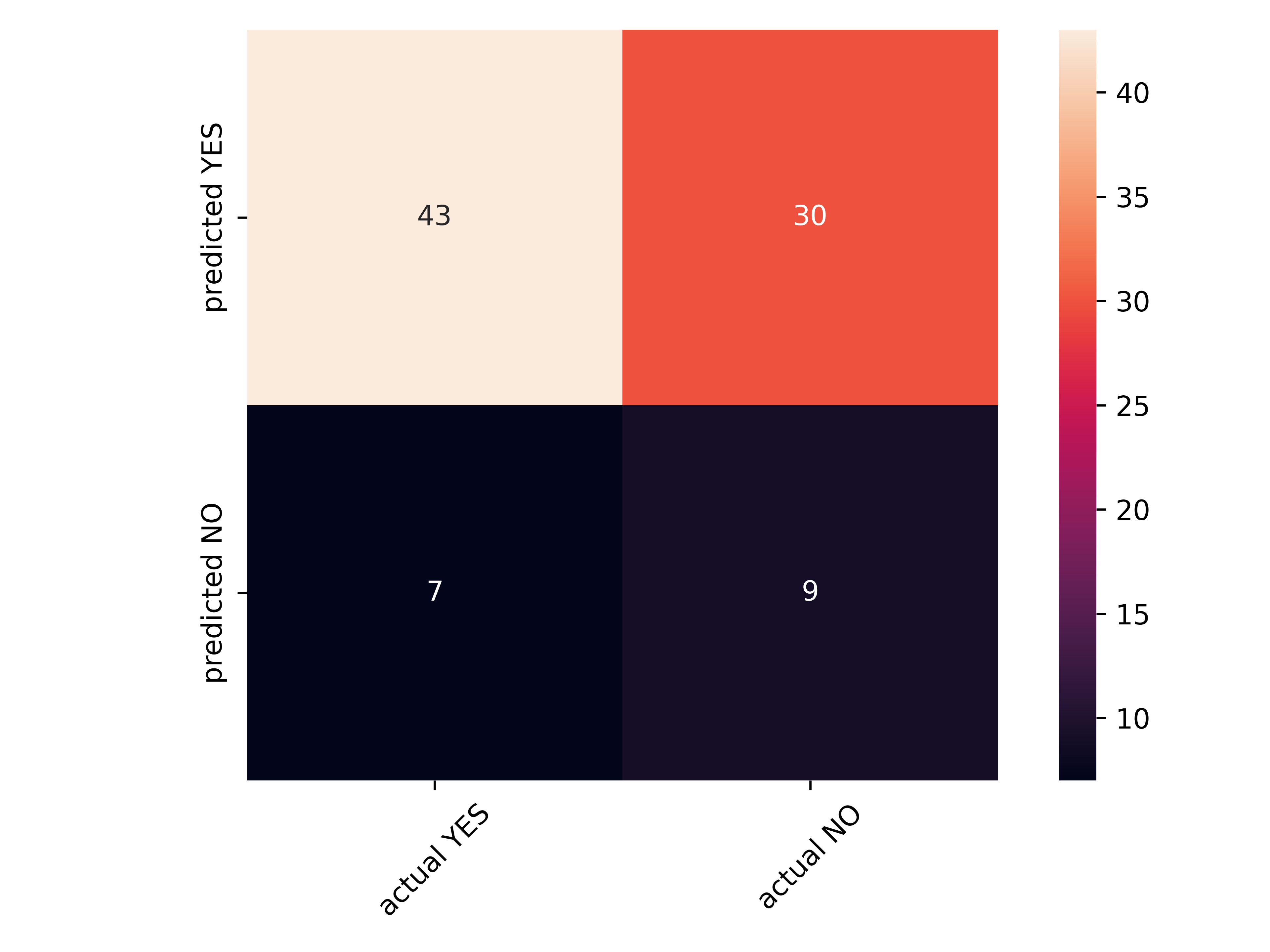}
    \caption{Confusion matrices for \texttt{mixtral-8x7b-32768} correctness evaluation of: \texttt{gemma-7b-it} (top left), \texttt{llama3-8b-8192} (top right),  \texttt{llama3-70b-8192} (bottom left), \texttt{mixtral-8x7b-32768} (bottom right).}
    \label{fig:conf_matrices_mixtral}
\end{figure}

We can note that the strategy of \texttt{gemma-7b-it} as evaluator model is to reply (almost) always YES. Conversely, \texttt{llama3-8b-8192} chooses to reply almost always NO. The other models, \texttt{llama3-70b-8192} and \texttt{mixtral-8x7b-32768}, tend to reply YES more than NO, but confusion matrices for these models, shown in Figure~\ref{fig:conf_matrices_llama3_70b} and Figure~\ref{fig:conf_matrices_mixtral} respectively, reveal why precision, recall, and F1-score are also poor.


\subsubsection{Safety}
\begin{table*}[!ht]
    \caption{Safety self-evaluation results for each evaluator model, referring to a model to evaluate.
}
\label{tab:self_safety}
\centering
\setlength{\tabcolsep}{1mm} 
\begin{tabular}{llccccc}
   \toprule
   \textbf{Evaluator Model} & \textbf{Evaluated model} & \textbf{Detected vulnerabilities} & \textbf{Accuracy} & \textbf{Precision} & \textbf{Recall} & \textbf{F1} \\
   \midrule
   \texttt{gemma-7b-it} 
   & \texttt{gemma-7b-it} & 3/4 & 27\% & 1\% & 75\% & 2\%\\
   & \texttt{llama3-8b-8192} & 4/4 & 28\% & 1\% & 100\% & 3\%\\
   & \texttt{llama3-70b-8192} & 6/6 & 32\% & 2\% & 100\% & 5\%\\
   & \texttt{mixtral-8x7b-32768} & 10/11 & 30\% & 4\% & 91\% & 7\%\\
   \midrule
   \texttt{llama3-8b-8192} 
   & \texttt{gemma-7b-it} & 0/4 & 99\% & 0\% & 0\% & \#\\
   & \texttt{llama3-8b-8192} & 0/4 & 98\% & 0\% & 0\% & \#\\
   & \texttt{llama3-70b-8192} & 1/6 & 98\% & 100\% & 17\% & 29\%\\
   & \texttt{mixtral-8x7b-32768} & 0/11 & 97\% & \# & 0\% & \#\\
   \midrule
   \texttt{llama3-70b-8192} 
   & \texttt{gemma-7b-it} & 1/4 & 96\% & 8\% & 25\% & 13\%\\
   & \texttt{llama3-8b-8192} & 0/4 & 98\% & 0\% & 0\% & \#\\
   & \texttt{llama3-70b-8192} & 1/6 & 98\% & 25\% & 17\% & 20\%\\
   & \texttt{mixtral-8x7b-32768} & 3/11 & 96\% & 33\% & 27\% & 30\%\\
   \midrule
   \texttt{mixtral-8x7b-32768} 
   & \texttt{gemma-7b-it} & 1/4 & 87\% & 2\% & 25\% & 4\%\\
   & \texttt{llama3-8b-8192} & 1/4 & 82\% & 1\% & 20\% & 3\%\\
   & \texttt{llama3-70b-8192} & 5/6 & 83\% & 8\% & 71\% & 14\%\\
   & \texttt{mixtral-8x7b-32768} & 7/11 & 83\% & 13\% & 67\% & 21\%\\
   \bottomrule
\end{tabular}
\end{table*}
Vulnerabilities detection is a binary classification task, where models are asked to reply only YES or NO to the input.
Since the vulnerability analysis performed by Infer detected four types of vulnerabilities, there are 4 queries for each of the 356 GEN samples, totalling 1424 prompts per evaluator model. 

Table~\ref{tab:self_safety} shows the number of correctly detected vulnerabilities (true positives over vulnerabilities detected by Infer), and the overall accuracy for each evaluator model. 
Note that the total number of true positives here is 25 instead of 39, because multiple instances of the same vulnerability in a single file are conflated into a count of 1, as the classification task is binary.
Taking into account both measures, the detection of vulnerability is very low even if the overall accuracy is high; this is due to the fact that the ground truth is strongly unbalanced. To investigate further, we calculated precision, recall, and F1-score for each evaluated model. 
When one of the measures is not calculable, a \# mark is posed. For example, in three cases out of four, \texttt{llama3-8b-8192} has no true positives and no false positives, so the denominator of Precision is zero.
It is worth noting that when there are no true positives (no real vulnerabilities is detected), it is not possible to calculate the F1-score. 
Thus, a model that replies almost always NO gains a great accuracy, but it fails to detect those few vulnerabilities that are actually present. On the other extreme, \texttt{gemma-7b-it} behaves like in the self-correctness analysis, replying YES most of the times, succeeding in the detection of all vulnerabilities, but achieving a very low accuracy due to false positives.
All F1-scores are very low, under the value of 0.3.
In general, all evaluator LLMs do not seem to understand the vulnerabilities, even if the prompted code was generated by themselves.

\subsection{Repair}
\subsubsection{Correctness}
We asked the models to repair the 155 files (last three columns of Table~\ref{tab:correctness_results}) that were previously incorrect, or
for which some errors or timeouts occurred. 
Note that this includes incorrect files generated both by itself and by all the other models (without knowing which model generated the code, or how).


\begin{table}[!ht]
    \caption{Code correctness after the repair and code cleaning phase for each model.
\textbf{OK}: all the test passed,
\textbf{Exec}: an execution error or timeout occurred,
\textbf{Assert}: at least one test failed,
\textbf{Comp}: a compilation error occurred,
\textbf{\%}: percentage of correct files with respect to the 155 incorrect files.}
\label{tab:overall-correctness}
\centering
\setlength{\tabcolsep}{1mm} 
\renewcommand{\arraystretch}{1.1} 
\begin{tabular}{lccccc}
\toprule
\textbf{Model} & \textbf{OK} & \textbf{Exec} & \textbf{Assert} & \textbf{Comp} & \textbf{\%} \\
\midrule
\texttt{gemma-7b-it} & 20 & 19 & 91 & 4 & 13\% \\
\texttt{llama3-8b-8192} & 63 & 10 & 76 & 1 & 41\% \\
\texttt{llama3-70b-8192} & 92 & 8 & 52 & 2 & 59\% \\
\texttt{mixtral-8x7b-32768} & 51 & 14 & 71 & 2 & 33\% \\
\midrule
\textbf{Overall} & 226 & 51 & 290 & 9 & ~ \\
\bottomrule
\end{tabular} 
\end{table}

\begin{table}[!ht]
    \caption{Code correctness in terms of number of files after the repair and code cleaning phase for different models.
    \textbf{Repair}: model that performs the repair phase,
    \textbf{Generated by}: model that generated the code to repair,
   \textbf{OK}: all the test passed,
    \textbf{Exec}: an execution error or timeout occurred,
    \textbf{Assert}: at least one test failed,
    \textbf{Comp}: a compilation error occurred.
}
\label{tab:aggregated_correctness}
\centering
\setlength{\tabcolsep}{0.5mm} 
\begin{tabular}{llcccc}
    \toprule
    \textbf{Repair} & \textbf{Generated by} & \textbf{OK} & \textbf{Exec} & \textbf{Assert} & \textbf{Comp} \\
    \midrule
    \texttt{gemma-7b-it} & \texttt{gemma-7b-it} & 5 & 5 & 31 & 2 \\
    & \texttt{llama3-8b-8192} & 9 & 6 & 15 & 0 \\
    & \texttt{llama3-70b-8192} & 3 & 2 & 22 & 0 \\
    & \texttt{mixtral-8x7b-32768} & 3 & 6 & 23 & 2 \\
    \midrule
    \texttt{llama3-8b-8192} & \texttt{gemma-7b-it} & 26 & 3 & 17 & 0 \\
    & \texttt{llama3-8b-8192} & 11 & 3 & 23 & 0 \\
    & \texttt{llama3-70b-8192} & 13 & 2 & 13 & 1 \\
    & \texttt{mixtral-8x7b-32768} & 13 & 2 & 23 & 0 \\
    \midrule
    \texttt{llama3-70b-8192} & \texttt{gemma-7b-it} & 33 & 2 & 13 & 0 \\
    & \texttt{llama3-8b-8192} & 20 & 2 & 14 & 1 \\
    & \texttt{llama3-70b-8192} & 16 & 2 & 13 & 0 \\
    & \texttt{mixtral-8x7b-32768} & 23 & 2 & 12 & 1 \\
    \midrule
    \texttt{mixtral-8x7b-} & \texttt{gemma-7b-it} & 19 & 4 & 21 & 0 \\
    \texttt{32768} & \texttt{llama3-8b-8192} & 12 & 4 & 17 & 0 \\
    & \texttt{llama3-70b-8192} & 10 & 2 & 14 & 1 \\
    & \texttt{mixtral-8x7b-32768} & 10 & 4 & 19 & 1 \\
    \bottomrule
\end{tabular}
\end{table}

\begin{figure}[!ht]
    \centering
    \begin{tabular}{cc}
        \begin{subfigure}[b]{0.49\columnwidth}
            \includegraphics[width=\textwidth]{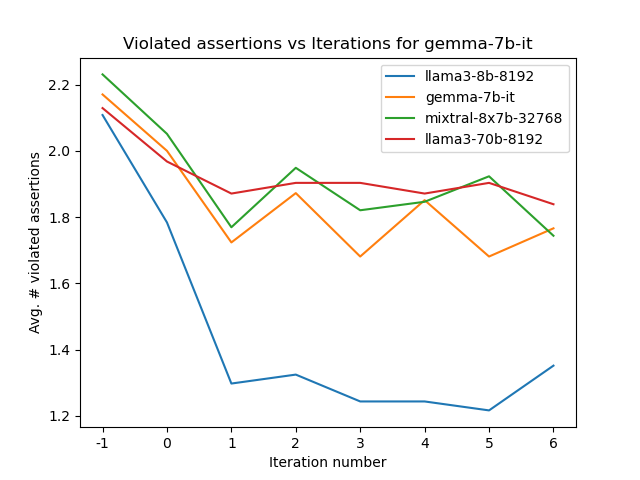}
            \caption{\texttt{gemma-7b-it}}
            \label{fig:gemma-7b-it-c}
        \end{subfigure} &
        \begin{subfigure}[b]{0.49\columnwidth}
            \includegraphics[width=\textwidth]{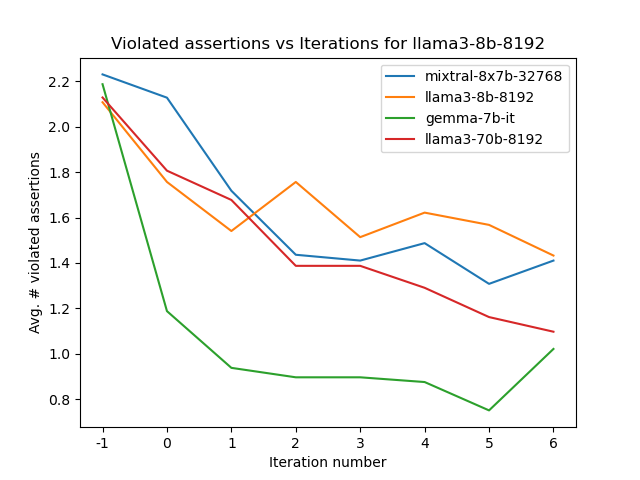}
            \caption{\texttt{llama3-8b-8192}}
            \label{fig:llama3-8b-8192-c}
        \end{subfigure} \\
        \begin{subfigure}[b]{0.49\columnwidth}
            \includegraphics[width=\textwidth]{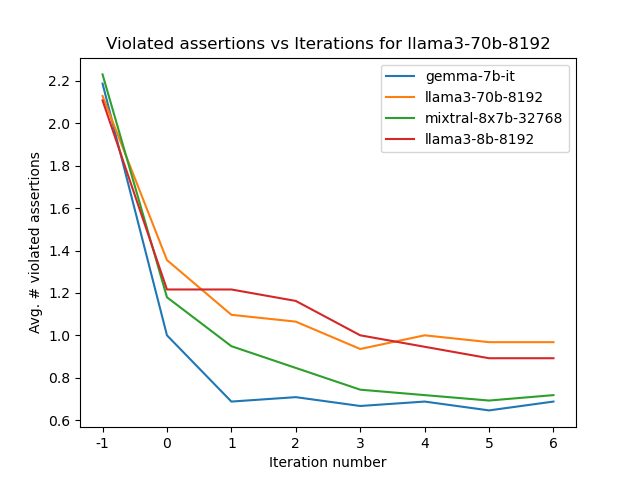}
            \caption{\texttt{llama3-70b-8192}}
            \label{fig:llama3-70b-8192-c}
        \end{subfigure} &
        \begin{subfigure}[b]{0.49\columnwidth}
            \includegraphics[width=\textwidth]{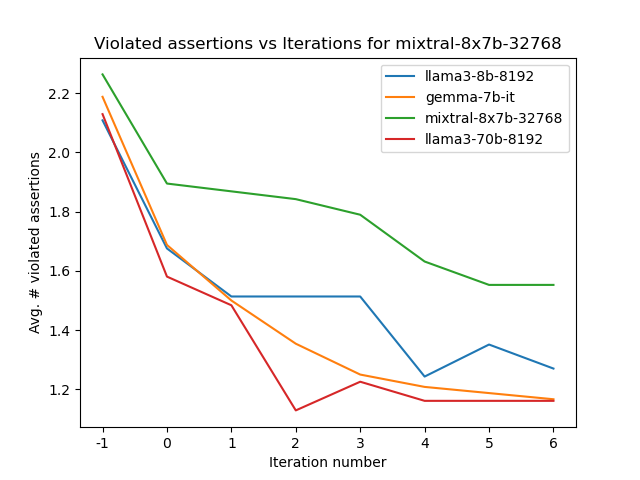}
            \caption{\texttt{mixtral-8x7b-32768}}
            \label{fig:mixtral-8x7b-32768-c}
        \end{subfigure}
    \end{tabular}
    \caption{Correctness analysis results for different models in which the average number of test failed for each
    iteration is shown.
    Iteration -1 refers to the number of assertion failures in the original code, before
    the repair phase.
    }
    \label{fig:correctness-iter-analysis}
\end{figure}

We analyzed the supposedly repaired files using the same correctness
analysis pipeline as before, and the results are shown in
Tables~\ref{tab:overall-correctness} and ~\ref{tab:aggregated_correctness}.
We can see how \texttt{llama3-70b-8192} fixes the highest number of incorrect
files, while \texttt{gemma-7b-it} fixes the lowest one.
When analyzing the results for each model performing the
repair phase, as reported in Table~\ref{tab:aggregated_correctness},
we can note that there is no emergent trend
of self-preference (i.e., a model that fixes better the code generated by itself).
On the contrary, for all the models except \texttt{gemma-7b-it},
the percentage of corrected files is lower
when trying to repair the code generated by themselves, ranging from 51\% to 25\%.
Except for \texttt{gemma-7b-it}, all the other models, on average,
fix more files generated by \texttt{gemma-7b-it} than by themselves or other models.
The model that performs better is
\texttt{llama3-70b-8192}, which fixes 68\% of the incorrect files
for \texttt{gemma-7b-it} and 59\% if we consider the average for all the models.
No model is able to fix all the incorrect files.

Figure~\ref{fig:correctness-iter-analysis} shows the average number of failed tests
for each iteration and for each model. We can see how \texttt{llama3-70b-8192} has the steepest decrease in the number of failed tests, and it is also the one in which the plotted lines are lower and more stable after just two iterations.

\subsubsection{Safety}
\begin{table}[!t]
    \caption{Aggregate vulnerability analysis for each model after the repair phase.
    \textbf{-}: number of removed vulnerabilities
    (i.e., fixed) by the
    model that performed the repair phase, \textbf{=}: number of vulnerabilities still present after the repair phase,
    \textbf{+}: number of vulnerabilities added
    after the repair phase (i.e., the ones that were not present in the initial analysis), \textbf{\%}: percentage of fixed vulnerabilities.
}
\label{tab:infer_bugs}
\centering
\setlength{\tabcolsep}{1mm} 
\renewcommand{\arraystretch}{1.1} 
\begin{tabular}{lcccc}
    \toprule
    \textbf{Model} & \textbf{-} & \textbf{=} & \textbf{+} & \textbf{\%} \\
    \midrule
    \texttt{gemma-7b-it} & 24 & 14 & 1 & 63\% \\
    \texttt{llama3-8b-8192} & 29 & 15 & 2 & 66\% \\
    \texttt{llama3-70b-8192} & 39 & 5 & 0 & 89\% \\
    \texttt{mixtral-8x7b-32768} & 39 & 5 & 0 & 89\% \\
    \midrule
    \textbf{Cumulative} & 131 & 39 & 3 \\
    \bottomrule
\end{tabular}
\end{table}
\begin{table}[!t]
    \caption{Vulnerability analysis for each model after the repair phase.
    \textbf{Repair}: model that performs the repair phase,
    \textbf{Generated by}: model that generated the code to repair,
    \textbf{-}: number of removed vulnerabilities,
    \textbf{=}: number of remaining vulnerabilities,
    \textbf{+}: number of added vulnerabilities.
}
\label{tab:aggregated_bugs}
\centering
\setlength{\tabcolsep}{1.5mm} 
\begin{tabular}{llccc}
    \toprule
    \textbf{Repair} & \textbf{Generated by} & \textbf{-} & \textbf{=} & \textbf{+} \\
    \midrule
    \texttt{gemma-7b-it} & \texttt{gemma-7b-it} & 4 & 3 & 0 \\
    & \texttt{llama3-8b-8192} & 5 & 2 & 0 \\
    & \texttt{llama3-70b-8192} & 9 & 1 & 1 \\
    & \texttt{mixtral-8x7b-32768} & 6 & 8 & 0 \\
    \midrule
    \texttt{llama3-8b-8192} & \texttt{gemma-7b-it} & 7 & 0 & 0 \\
    & \texttt{llama3-8b-8192} & 4 & 4 & 0 \\
    & \texttt{llama3-70b-8192} & 10 & 0 & 0 \\
    & \texttt{mixtral-8x7b-32768} & 8 & 11 & 2 \\
    \midrule
    \texttt{llama3-70b-8192} & \texttt{gemma-7b-it} & 7 & 0 & 0 \\
    & \texttt{llama3-8b-8192} & 8 & 0 & 0 \\
    & \texttt{llama3-70b-8192} & 10 & 0 & 0 \\
    & \texttt{mixtral-8x7b-32768} & 14 & 5 & 0 \\
    \midrule
    \texttt{mixtral-8x7b-32768} & \texttt{gemma-7b-it} & 7 & 0 & 0 \\
    & \texttt{llama3-8b-8192} & 8 & 0 & 0 \\
    & \texttt{llama3-70b-8192} & 9 & 1 & 0 \\
    & \texttt{mixtral-8x7b-32768} & 15 & 4 & 0 \\
    \bottomrule
\end{tabular}
\end{table}
We asked the models to repair files for which Infer had detected at least one vulnerability, namely the ones reported in the last column of Table~\ref{tab:safe_unsafe_files}.
%
This includes the false positives reported by Infer, to reflect a practical application of this approach, where the static analyser is run as part of a code generation pipeline without a human in the loop to verify each issue.
Note that each model is accountable for fixing the vulnerabilities in the code generated by itself and by all the other models. 

On the supposedly repaired files, we run Infer with the same configuration as before, checking if vulnerabilities were actually fixed.
Table~\ref{tab:infer_bugs} summarizes the vulnerability analysis on the regenerated files, reporting, for each model, the number of removed, remaining, and newly added vulnerabilities.
The total number of vulnerabilities (sum of ``-'' and ``='' columns) is lower for \texttt{gemma-7b-it} due to compilation errors that is introduced during the regeneration of samples previously generated by \texttt{llama3-8b-8192} and \texttt{mixtral-8x7b-32768} (see Table~\ref{tab:aggregated_bugs}).

Overall, $\sim$77\% of the
previously detected vulnerabilities has been fixed by the models, while 39 out of
the original 176 (44 vulnerabilities for 4 models) vulnerabilities are still detected
by Infer. 
By manually inspecting all these vulnerabilities, we found that just one
vulnerability was a true positive, while the others are all false positives. 

Note that 3 newly vulnerabilities were detected by Infer on the new generated files
(last column of Table~\ref{tab:infer_bugs}), but after manual inspection, we found
that they are all false positives.

Both \texttt{llama3-70b-8192} and \texttt{mixtral-8x7b-32768}
have the same number of removed vulnerabilities and they do not have any added
vulnerabilities. Interestingly, the models also fixed some vulnerabilities that,
after manual inspection, were found to be false positives.

\begin{figure}[t]
	\centering
	\begin{tabular}{cc}
		\begin{subfigure}[b]{0.49\columnwidth}
			\includegraphics[width=\textwidth]{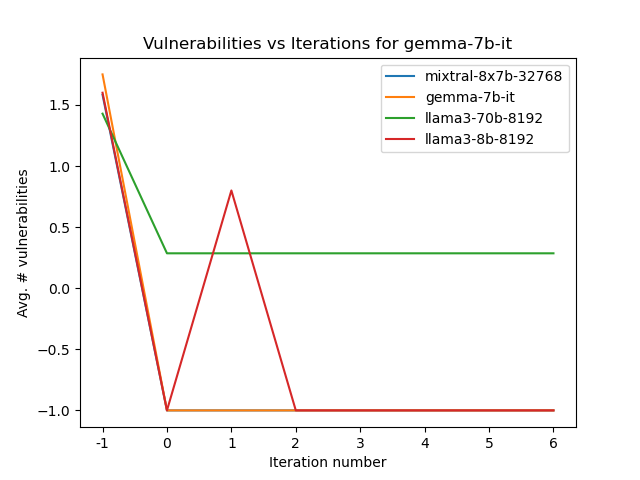}
			\caption{\texttt{gemma-7b-it}}
			\label{fig:gemma-7b-it-v}
		\end{subfigure} &
		\begin{subfigure}[b]{0.49\columnwidth}
			\includegraphics[width=\textwidth]{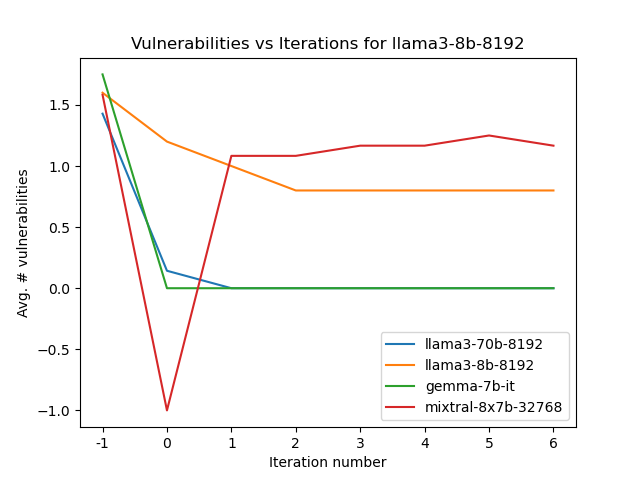}
			\caption{\texttt{llama3-8b-8192}}
			\label{fig:llama3-8b-8192-v}
		\end{subfigure} \\
		\begin{subfigure}[b]{0.49\columnwidth}
			\includegraphics[width=\textwidth]{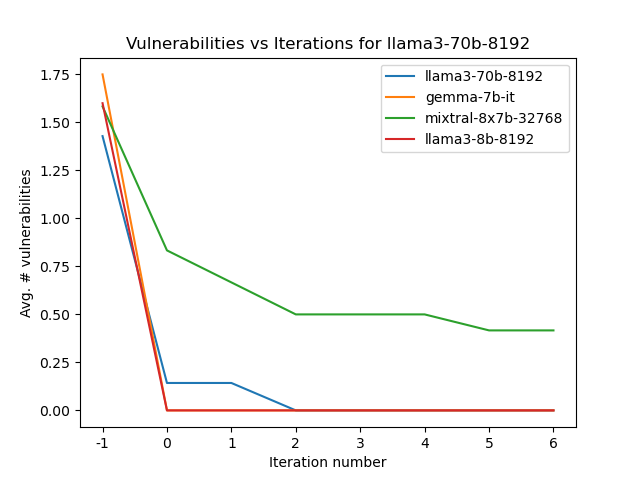}
			\caption{\texttt{llama3-70b-8192}}
			\label{fig:llama3-70b-8192-v}
		\end{subfigure} &
		\begin{subfigure}[b]{0.49\columnwidth}
			\includegraphics[width=\textwidth]{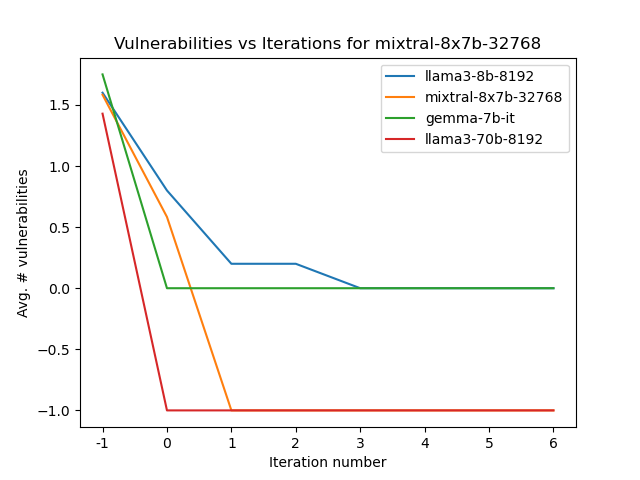}
			\caption{\texttt{mixtral-8x7b-32768}}
			\label{fig:mixtral-8x7b-32768-v}
		\end{subfigure}
	\end{tabular}
	\caption{Vulnerability analysis results for each model showing the average
		number of vulnerabilities found for each iteration. The value -1 means that
		the model produce code that did not compile after that iteration. Iteration -1 refers to the number of vulnerabilities in the original code, before
		the repair phase.}
	\label{fig:vuln_analysis}
\end{figure}

We present a breakdown of the vulnerability analysis for each model in Table~\ref{tab:aggregated_bugs}.
Similarly to the correctness analysis, also for the safety analysis we can see that there is no emergent trend
of self-preference.
Except for \texttt{gemma-7b-it}, the other models are able to fix all the vulnerabilities generated by two other models (3 in the case of \texttt{llama3-70b-8192}).
Moreover, both \texttt{llama3-70b-8192} and \texttt{mixtral-8x7b-32768} fix
the vulnerabilities without introducing new ones.

For all the models, the highest percentage of success in fixing the
vulnerabilities is obtained when repairing the code generated by
\texttt{llama3-70b-8192}, with percentages ranging from 90\% to 100\%. Figure~\ref{fig:vuln_analysis} depicts the average number of vulnerabilities found
for each iteration for each model. Here it is possible to note how some models, after the first iteration, produce code that does not compile and therefore Infer is not able to analyze it,
some models though are able to recover from this situation and produce code that can be analyzed by Infer
and also fixing some vulnerabilities.

\section{Related Work}\label{sec:rw}

Due to the increasing adoption of LLM to perform coding tasks, new
studies regarding the quality and safety evaluation of AI-generated code have rapidly emerged.
Some studies concern only the generation and assessment of AI-generated code, 
\rebut{
    and focus on the evaluation of CWE vulnerabilities.
One of these is} the FormAI Dataset~\cite{TihanyiBJFCM23} that performed vulnerability evaluation using
Efficient SMT-based Bounded Model Checker (ESBMC)~\cite{GadelhaMMC0N18} on
112000 AI-generated C programs using a dynamic zero-shot prompting technique,
revealing that 51.24\% of the programs contain vulnerabilities,
posing significant risks to software security.
\rebut{
Pearce et al.~\cite{r1_4_r3_5_2022copilot} examined cybersecurity weaknesses, focusing on Copilot. Their study reveals that Copilot introduces vulnerabilities in several cases, demonstrating that it is not yet capable of guaranteeing the security of the generated code.
Additionally, experimenting with various strategies to improve Copilot's security, none of these proved to be entirely effective in removing vulnerabilities.

To further explore LLMs performance, other works focus on code repair scenarios~\cite{abs-2305-14752,JinSTSLSS23,abs-2311-02433,Li0ZQ23,r3_1_2024livecodebench,r3_2_2023selfrepair,r4_2_2024enhancing,r3_3_Ullah2024}.}
\cite{abs-2305-14752} measures the
self-healing capabilities of LLMs using ESBMC to
identify vulnerabilities and produce counterexamples, which are then fed to
gpt-3.5-turbo~\cite{gpt35turbo}, showing a high success rate in repairing
vulnerabilities like buffer overflow and pointer dereference failures.
Frameworks for program repair based on static analyzers like InferFix~\cite{JinSTSLSS23}
employ LLMs fine-tuned on bug repair patches to solve core repairing tasks
for three types of bugs.
Other studies tried to show the possible outcomes of integrating LLMs into
static analysis tasks, such as loop invariant annotations~\cite{abs-2311-02433}
and pruning false positive in bug detections~\cite{Li0ZQ23}.

While works~\cite{r1_4_r3_5_2022copilot,abs-2305-14752,JinSTSLSS23,abs-2311-02433,Li0ZQ23} mentioned above focus on GPTs models, which are not entirely
open-source and require users to pay for using the inference or training API,
our experimental study only relies on open-source models and requires
no fine-tuning.
Furthermore, these methodologies do not check for the \rebut{functional} correctness of the generated code with respect to the specifications (i.e., the user prompts), nor measure the extent of self-awareness (or lack of--) of LLMs when faced with their own generated code. Our work focuses on correctness and safety, by evaluating both dimensions with tests, analysis, and by asking to the model.

In recent works, such as~\cite{ChapmanSOAP2024} and~\cite{Li2024}, 
novel methodologies to use LLMs for static analysis were proposed, or rather to use LLMs \emph{as} static analyzers.
The key idea of~\cite{ChapmanSOAP2024} is to interleave the static analyzer reports with the LLMs' responses, iterating over a single piece of code.
\cite{Li2024}, instead, described a neuro-symbolic framework to perform vulnerability detection for a whole repository.
Although the purposes of these works are very different from ours, the intertwining of LLMs and static analyzers proved to be effective, thus we also employ an iterative method in the repair phase.

\rebut{
    Other works rely solely on LLMs' abilities for code self-repair~\cite{r3_1_2024livecodebench,r3_2_2023selfrepair,r4_2_2024enhancing,r3_3_Ullah2024}.
    In~\cite{r3_1_2024livecodebench}, LLMs are tasked with repairing code that yields incorrect results based on their dataset test cases, and do not actively search for safety vulnerabilities in the code. 
    Both \cite{r3_1_2024livecodebench} and~\cite{r3_2_2023selfrepair} lack an iterative repair process, which is integral to our framework for progressively improving LLM outputs.
    \cite{r4_2_2024enhancing} proposes a novel LLM that acts as an automated security code repair system. Notably, we did not train nor fine-tune our chosen models, while the proposal \cite{r4_2_2024enhancing} performed supervised fine-tuning and reinforcement learning to train a specialized LLM. \cite{r3_3_Ullah2024}  introduces the SecLLMHolmes framework, that relies on specific prompts and the security reasoning capabilities of LLMs to detect potential issues.
}
Regarding LLMs' self-awareness, in~\cite{r3_3_Ullah2024} authors performed an extensive evaluation on how well LLMs could detect bugs and security-related issues in code, finding that they perform poorly and give incorrect responses. 
In our work, we also found that LLMs have a lack of understanding of the possible vulnerabilities in their own generated code. As an addition, we also tested LLMs capabilities in judging the correctness of their own code, and analyzed the self-preferences, which was out of the scope of~\cite{r3_3_Ullah2024}. \rebut{
Many of the cited works focus on cybersecurity-related issues, while we specifically target code safety. For example, \cite{r3_4_r4_1_2023zeroshot} investigates the ability of LLMs to generate and repair specific kinds of security vulnerabilities (e.g.,  CWE-787, CWE-89), applying a single zero-shot approach without iterative repair. Despite this, our results align with the conclusion of~\cite{r1_4_r3_5_2022copilot,r3_1_2024livecodebench,r3_2_2023selfrepair,r4_2_2024enhancing,r3_4_r4_1_2023zeroshot,r3_3_Ullah2024}, finding significant limitations in the usage of LLMs to detect vulnerabilities, and realizing that state-of-the-art models are not yet adequate to perform zero-shot code repair in an automated framework. We contribute to this key finding by proposing a completely automated and iterative process that guides LLMs to enhance their initial baseline of generated code. Our work boosts this aspect by means of the intervention of static analysis in the pipeline, providing accurate feedback and guiding LLMs toward an improvement of their own generated code.
}

\rebut{Regarding datasets, \cite{abs-2312-04724,HeVKV24} propose cybersecurity-focused datasets for evaluating LLMs' code generation. \cite{r3_1_2024livecodebench} presents LiveCodeBench, a benchmark for testing LLMs with datasets of general programming problems to address the limitations of benchmarks like MBPP. However, problems in LiveCodeBench are gathered from competitions and other sources, and the text of the task could vary widely among each other. Instead, our tailored version of MBPP for C asks for a single function, of which we know the whole specification.  While complementary to our work, we prioritized both the correctness and safety of the generated code, with security being outside
the primary scope of this~paper.}

\section{Conclusions}\label{sec:concl}

In this paper, we present an extensive evaluation of Large Language
Models (LLMs) in generating code that complies with user prompts,
evaluating (in)correctness, and addressing potential
vulnerabilities using static analysis, with the addition of a self-awareness
evaluation on these two topics. Our study involved the creation of a
benchmark from the Mostly Basic Python Problems Dataset (MBPP)
adapted to the C language.

After the generation phase, we analyzed the compiling
programs using the Infer static analyzer and we ran tests to find
potential vulnerabilities and errors. Using this feedback, we evaluated the
ability of LLMs to repair code regarding both safety and
correctness.

Our experiments revealed a gap in LLMs' ability to generate fully
compliant and secure code autonomously. The static analysis
highlighted some vulnerabilities, underscoring the need to integrate
robust verification and validation steps into the code generation
process. Furthermore, our findings indicate that LLMs exhibit
a very low degree of self-awareness about the correctness and
safety of their generated code. On the other hand, when prompted
with issues, on both safety and correctness, these models
show some capabilities of repairing the code they or other models
have produced. 
Our results also emphasize the necessity for continued
advancements in LLMs' understanding and application of secure
coding practices. Moreover, automated pipelines like the one
we propose could help create safer and more reliable code generation.

The proposed pipeline is designed to be fully modular,
allowing for a rapid adaptation and easy interchangeability of considered LLMs,
static analyzers, and datasets. \rebut{A promising direction for future work would be to investigate how to integrate a combination of static analyzers, such as Infer and CodeQL,%
\footnote{\url{https://codeql.github.com/}}
to leverage their complementary strengths and reduce the risk of false positives and negatives.}

Concerning safety repair, our current approach is based on static analysis using Infer. Future works will explore hybrid approaches, for instance by combining static and dynamic analyses. The outputs from these mixed analyses will serve as feedback for LLMs, enhancing the reliability and safety of the code they generate. Finally, while this paper focuses on the C programming language, where particular attention must be paid to memory management, we think each programming language presents unique challenges for LLMs in generating reliable and safe code. We plan to extend our studies to other programming languages, such as Rust, where the strict requirements for memory safety offer an interesting context to evaluate the ability of LLMs to handle complex safety constraints, or Python, where it will be interesting to see how LLMs perform on a dynamically-typed and highly-flexible language.
In future work, we also plan to evaluate whether the generated code is indeed newly generated code or simply the translation of code that the model has been inadvertently trained on.

\section{Data Availability}
We provide\footnote{\url{https://figshare.com/s/4a2e799b97bfeacea0c3}. For the submission phase, we provide an anonymous link to our data. If accepted, we will move the data to a Zenodo repository with DOI.} (i) the benchmark suite of tasks (Section~\ref{sec:expset}), including both the raw and cleaned source code generated by the models we considered, (ii) the source code for the scripts used in our experimental evaluation (Section~\ref{sec:eval}), (iii) the analysis results computed by Infer, covering both the vulnerability analysis and the repair phases.

%


\clearpage
\bibliography{aaai25}

\begin{thebibliography}{10}
\providecommand{\url}[1]{#1}
\csname url@samestyle\endcsname
\providecommand{\newblock}{\relax}
\providecommand{\bibinfo}[2]{#2}
\providecommand{\BIBentrySTDinterwordspacing}{\spaceskip=0pt\relax}
\providecommand{\BIBentryALTinterwordstretchfactor}{4}
\providecommand{\BIBentryALTinterwordspacing}{\spaceskip=\fontdimen2\font plus
\BIBentryALTinterwordstretchfactor\fontdimen3\font minus
  \fontdimen4\font\relax}
\providecommand{\BIBforeignlanguage}[2]{{%
\expandafter\ifx\csname l@#1\endcsname\relax
\typeout{** WARNING: IEEEtran.bst: No hyphenation pattern has been}%
\typeout{** loaded for the language `#1'. Using the pattern for}%
\typeout{** the default language instead.}%
\else
\language=\csname l@#1\endcsname
\fi
#2}}
\providecommand{\BIBdecl}{\relax}
\BIBdecl

\bibitem{abs-2401-05949}
\BIBentryALTinterwordspacing
S.~Zhao, M.~Jia, L.~A. Tuan, and J.~Wen, ``Universal vulnerabilities in large
  language models: In-context learning backdoor attacks,'' \emph{CoRR}, vol.
  abs/2401.05949, 2024. [Online]. Available:
  \url{https://doi.org/10.48550/arXiv.2401.05949}
\BIBentrySTDinterwordspacing

\bibitem{Zhang0ZAW23}
\BIBentryALTinterwordspacing
B.~Zhang, P.~Liang, X.~Zhou, A.~Ahmad, and M.~Waseem, ``Practices and
  challenges of using github copilot: An empirical study,'' in \emph{The 35th
  International Conference on Software Engineering and Knowledge Engineering,
  {SEKE} 2023, {KSIR} Virtual Conference Center, USA, July 1-10, 2023},
  S.~Chang, Ed.\hskip 1em plus 0.5em minus 0.4em\relax {KSI} Research Inc.,
  2023, pp. 124--129. [Online]. Available:
  \url{https://doi.org/10.18293/SEKE2023-077}
\BIBentrySTDinterwordspacing

\bibitem{BarkeJP23}
\BIBentryALTinterwordspacing
S.~Barke, M.~B. James, and N.~Polikarpova, ``Grounded copilot: How programmers
  interact with code-generating models,'' \emph{Proc. {ACM} Program. Lang.},
  vol.~7, no. {OOPSLA1}, pp. 85--111, 2023. [Online]. Available:
  \url{https://doi.org/10.1145/3586030}
\BIBentrySTDinterwordspacing

\bibitem{gartner}
\BIBentryALTinterwordspacing
Gartner, ``Gartner hype cycle shows ai practices and platform engineering will
  reach mainstream adoption in software engineering in two to five years,''
  2024. [Online]. Available:
  \url{\url{https://www.gartner.com/en/newsroom/press-releases/2023-11-28-gartner-hype-cycle-shows-ai-practices-and-platform-engineering-will-reach-mainstream-adoption-in-software-engineering-in-two-to-five-years}}
\BIBentrySTDinterwordspacing

\bibitem{llama3}
\BIBentryALTinterwordspacing
M.~L. Team, ``https://ai.meta.com/blog/meta-llama-3/,'' 2024. [Online].
  Available: \url{\url{https://ai.meta.com/blog/meta-llama-3/}}
\BIBentrySTDinterwordspacing

\bibitem{abs-2403-08295}
T.~Mesnard, C.~Hardin, R.~Dadashi, S.~Bhupatiraju, S.~Pathak, L.~Sifre,
  M.~Rivi{\`{e}}re, M.~S. Kale, J.~Love, P.~Tafti, L.~Hussenot, A.~Chowdhery,
  A.~Roberts, A.~Barua, A.~Botev, A.~Castro{-}Ros, A.~Slone, A.~H{\'{e}}liou,
  A.~Tacchetti, A.~Bulanova, A.~Paterson, B.~Tsai, B.~Shahriari, C.~L. Lan,
  C.~A. Choquette{-}Choo, C.~Crepy, D.~Cer, D.~Ippolito, D.~Reid,
  E.~Buchatskaya, E.~Ni, E.~Noland, G.~Yan, G.~Tucker, G.~Muraru,
  G.~Rozhdestvenskiy, H.~Michalewski, I.~Tenney, I.~Grishchenko, J.~Austin,
  J.~Keeling, J.~Labanowski, J.~Lespiau, J.~Stanway, J.~Brennan, J.~Chen,
  J.~Ferret, J.~Chiu, and et~al., ``Gemma: Open models based on gemini research
  and technology,'' \emph{CoRR}, vol. abs/2403.08295, 2024.

\bibitem{abs-2401-04088}
\BIBentryALTinterwordspacing
A.~Q. Jiang, A.~Sablayrolles, A.~Roux, A.~Mensch, B.~Savary, C.~Bamford, D.~S.
  Chaplot, D.~de~Las~Casas, E.~B. Hanna, F.~Bressand, G.~Lengyel, G.~Bour,
  G.~Lample, L.~R. Lavaud, L.~Saulnier, M.~Lachaux, P.~Stock, S.~Subramanian,
  S.~Yang, S.~Antoniak, T.~L. Scao, T.~Gervet, T.~Lavril, T.~Wang, T.~Lacroix,
  and W.~E. Sayed, ``Mixtral of experts,'' \emph{CoRR}, vol. abs/2401.04088,
  2024. [Online]. Available: \url{https://doi.org/10.48550/arXiv.2401.04088}
\BIBentrySTDinterwordspacing

\bibitem{CalcagnoD11}
\BIBentryALTinterwordspacing
C.~Calcagno and D.~Distefano, ``Infer: An automatic program verifier for memory
  safety of {C} programs,'' in \emph{{NASA} Formal Methods - Third
  International Symposium, {NFM} 2011, Pasadena, CA, USA, April 18-20, 2011.
  Proceedings}, ser. Lecture Notes in Computer Science, M.~G. Bobaru,
  K.~Havelund, G.~J. Holzmann, and R.~Joshi, Eds., vol. 6617.\hskip 1em plus
  0.5em minus 0.4em\relax Springer, 2011, pp. 459--465. [Online]. Available:
  \url{https://doi.org/10.1007/978-3-642-20398-5\_33}
\BIBentrySTDinterwordspacing

\bibitem{chen2021codex}
M.~Chen, J.~Tworek, H.~Jun, Q.~Yuan, H.~P. de~Oliveira~Pinto, J.~Kaplan,
  H.~Edwards, Y.~Burda, N.~Joseph, G.~Brockman, A.~Ray, R.~Puri, G.~Krueger,
  M.~Petrov, H.~Khlaaf, G.~Sastry, P.~Mishkin, B.~Chan, S.~Gray, N.~Ryder,
  M.~Pavlov, A.~Power, L.~Kaiser, M.~Bavarian, C.~Winter, P.~Tillet, F.~P.
  Such, D.~Cummings, M.~Plappert, F.~Chantzis, E.~Barnes, A.~Herbert-Voss,
  W.~H. Guss, A.~Nichol, A.~Paino, N.~Tezak, J.~Tang, I.~Babuschkin, S.~Balaji,
  S.~Jain, W.~Saunders, C.~Hesse, A.~N. Carr, J.~Leike, J.~Achiam, V.~Misra,
  E.~Morikawa, A.~Radford, M.~Knight, M.~Brundage, M.~Murati, K.~Mayer,
  P.~Welinder, B.~McGrew, D.~Amodei, S.~McCandlish, I.~Sutskever, and
  W.~Zaremba, ``Evaluating large language models trained on code,'' 2021.

\bibitem{abs-2310-06825}
\BIBentryALTinterwordspacing
A.~Q. Jiang, A.~Sablayrolles, A.~Mensch, C.~Bamford, D.~S. Chaplot,
  D.~de~Las~Casas, F.~Bressand, G.~Lengyel, G.~Lample, L.~Saulnier, L.~R.
  Lavaud, M.~Lachaux, P.~Stock, T.~L. Scao, T.~Lavril, T.~Wang, T.~Lacroix, and
  W.~E. Sayed, ``Mistral 7b,'' \emph{CoRR}, vol. abs/2310.06825, 2023.
  [Online]. Available: \url{https://doi.org/10.48550/arXiv.2310.06825}
\BIBentrySTDinterwordspacing

\bibitem{abs-2406-06608}
S.~Schulhoff, M.~Ilie, N.~Balepur, K.~Kahadze, A.~Liu, C.~Si, Y.~Li, A.~Gupta,
  H.~Han, S.~Schulhoff, P.~S. Dulepet, S.~Vidyadhara, D.~Ki, S.~Agrawal,
  C.~Pham, G.~C. Kroiz, F.~Li, H.~Tao, A.~Srivastava, H.~D. Costa, S.~Gupta,
  M.~L. Rogers, I.~Goncearenco, G.~Sarli, I.~Galynker, D.~Peskoff, M.~Carpuat,
  J.~White, S.~Anadkat, A.~M. Hoyle, and P.~Resnik, ``The prompt report: {A}
  systematic survey of prompting techniques,'' \emph{CoRR}, vol.
  abs/2406.06608, 2024.

\bibitem{Wei0SBIXCLZ22}
\BIBentryALTinterwordspacing
J.~Wei, X.~Wang, D.~Schuurmans, M.~Bosma, B.~Ichter, F.~Xia, E.~H. Chi, Q.~V.
  Le, and D.~Zhou, ``Chain-of-thought prompting elicits reasoning in large
  language models,'' in \emph{Advances in Neural Information Processing Systems
  35: Annual Conference on Neural Information Processing Systems 2022, NeurIPS
  2022, New Orleans, LA, USA, November 28 - December 9, 2022}, S.~Koyejo,
  S.~Mohamed, A.~Agarwal, D.~Belgrave, K.~Cho, and A.~Oh, Eds., 2022. [Online].
  Available:
  \url{http://papers.nips.cc/paper\_files/paper/2022/hash/9d5609613524ecf4f15af0f7b31abca4-Abstract-Conference.html}
\BIBentrySTDinterwordspacing

\bibitem{ZhouMHPPCB23}
\BIBentryALTinterwordspacing
Y.~Zhou, A.~I. Muresanu, Z.~Han, K.~Paster, S.~Pitis, H.~Chan, and J.~Ba,
  ``Large language models are human-level prompt engineers,'' in \emph{The
  Eleventh International Conference on Learning Representations, {ICLR} 2023,
  Kigali, Rwanda, May 1-5, 2023}.\hskip 1em plus 0.5em minus 0.4em\relax
  OpenReview.net, 2023. [Online]. Available:
  \url{https://openreview.net/forum?id=92gvk82DE-}
\BIBentrySTDinterwordspacing

\bibitem{CousotC77}
\BIBentryALTinterwordspacing
P.~Cousot and R.~Cousot, ``Abstract interpretation: {A} unified lattice model
  for static analysis of programs by construction or approximation of
  fixpoints,'' in \emph{Conference Record of the Fourth {ACM} Symposium on
  Principles of Programming Languages, Los Angeles, California, USA, January
  1977}, R.~M. Graham, M.~A. Harrison, and R.~Sethi, Eds.\hskip 1em plus 0.5em
  minus 0.4em\relax {ACM}, 1977, pp. 238--252. [Online]. Available:
  \url{https://doi.org/10.1145/512950.512973}
\BIBentrySTDinterwordspacing

\bibitem{TihanyiBJFCM23}
N.~Tihanyi, T.~Bisztray, R.~Jain, M.~A. Ferrag, L.~C. Cordeiro, and
  V.~Mavroeidis, ``The formai dataset: Generative {AI} in software security
  through the lens of formal verification,'' in \emph{Proceedings of the 19th
  International Conference on Predictive Models and Data Analytics in Software
  Engineering, {PROMISE} 2023, San Francisco, CA, USA, 8 December 2023},
  S.~McIntosh, E.~Choi, and S.~Herbold, Eds.\hskip 1em plus 0.5em minus
  0.4em\relax {ACM}, 2023, pp. 33--43.

\bibitem{GadelhaMMC0N18}
\BIBentryALTinterwordspacing
M.~Y.~R. Gadelha, F.~R. Monteiro, J.~Morse, L.~C. Cordeiro, B.~Fischer, and
  D.~A. Nicole, ``{ESBMC} 5.0: an industrial-strength {C} model checker,'' in
  \emph{Proceedings of the 33rd {ACM/IEEE} International Conference on
  Automated Software Engineering, {ASE} 2018, Montpellier, France, September
  3-7, 2018}, M.~Huchard, C.~K{\"{a}}stner, and G.~Fraser, Eds.\hskip 1em plus
  0.5em minus 0.4em\relax {ACM}, 2018, pp. 888--891. [Online]. Available:
  \url{https://doi.org/10.1145/3238147.3240481}
\BIBentrySTDinterwordspacing

\bibitem{r1_4_r3_5_2022copilot}
H.~Pearce, B.~Ahmad, B.~Tan, B.~Dolan-Gavitt, and R.~Karri, ``Asleep at the
  keyboard? assessing the security of github copilot's code contributions,'' in
  \emph{IEEE Symposium on Security and Privacy, {S\&P} 2022}.\hskip 1em plus
  0.5em minus 0.4em\relax IEEE, 2022, pp. 754--768.

\bibitem{abs-2305-14752}
\BIBentryALTinterwordspacing
Y.~Charalambous, N.~Tihanyi, R.~Jain, Y.~Sun, M.~A. Ferrag, and L.~C. Cordeiro,
  ``A new era in software security: Towards self-healing software via large
  language models and formal verification,'' \emph{CoRR}, vol. abs/2305.14752,
  2023. [Online]. Available: \url{https://doi.org/10.48550/arXiv.2305.14752}
\BIBentrySTDinterwordspacing

\bibitem{JinSTSLSS23}
\BIBentryALTinterwordspacing
M.~Jin, S.~Shahriar, M.~Tufano, X.~Shi, S.~Lu, N.~Sundaresan, and
  A.~Svyatkovskiy, ``Inferfix: End-to-end program repair with llms,'' in
  \emph{Proceedings of the 31st {ACM} Joint European Software Engineering
  Conference and Symposium on the Foundations of Software Engineering,
  {ESEC/FSE} 2023, San Francisco, CA, USA, December 3-9, 2023}, S.~Chandra,
  K.~Blincoe, and P.~Tonella, Eds.\hskip 1em plus 0.5em minus 0.4em\relax
  {ACM}, 2023, pp. 1646--1656. [Online]. Available:
  \url{https://doi.org/10.1145/3611643.3613892}
\BIBentrySTDinterwordspacing

\bibitem{abs-2311-02433}
\BIBentryALTinterwordspacing
C.~Jan{\ss}en, C.~Richter, and H.~Wehrheim, ``Can chatgpt support software
  verification?'' \emph{CoRR}, vol. abs/2311.02433, 2023. [Online]. Available:
  \url{https://doi.org/10.48550/arXiv.2311.02433}
\BIBentrySTDinterwordspacing

\bibitem{Li0ZQ23}
\BIBentryALTinterwordspacing
H.~Li, Y.~Hao, Y.~Zhai, and Z.~Qian, ``Assisting static analysis with large
  language models: {A} chatgpt experiment,'' in \emph{Proceedings of the 31st
  {ACM} Joint European Software Engineering Conference and Symposium on the
  Foundations of Software Engineering, {ESEC/FSE} 2023, San Francisco, CA, USA,
  December 3-9, 2023}, S.~Chandra, K.~Blincoe, and P.~Tonella, Eds.\hskip 1em
  plus 0.5em minus 0.4em\relax {ACM}, 2023, pp. 2107--2111. [Online].
  Available: \url{https://doi.org/10.1145/3611643.3613078}
\BIBentrySTDinterwordspacing

\bibitem{r3_1_2024livecodebench}
N.~Jain, K.~Han, A.~Gu, W.-D. Li, F.~Yan, T.~Zhang, S.~Wang, A.~Solar-Lezama,
  K.~Sen, and I.~Stoica, ``Livecodebench: Holistic and contamination free
  evaluation of large language models for code,'' \emph{arXiv preprint
  arXiv:2403.07974}, 2024.

\bibitem{r3_2_2023selfrepair}
T.~X. Olausson, J.~P. Inala, C.~Wang, J.~Gao, and A.~Solar-Lezama, ``Is
  self-repair a silver bullet for code generation?'' in \emph{The Twelfth
  International Conference on Learning Representations}, 2023.

\bibitem{r4_2_2024enhancing}
N.~T. Islam, J.~Khoury, A.~Seong, E.~Bou-Harb, and P.~Najafirad, ``Enhancing
  source code security with llms: Demystifying the challenges and generating
  reliable repairs,'' in \emph{Network and Distributed System Security (NDSS)
  Symposium 2024}, 2024.

\bibitem{r3_3_Ullah2024}
S.~Ullah, M.~Han, S.~Pujar, H.~Pearce, A.~Coskun, and G.~Stringhini, ``{LLMs}
  cannot reliably identify and reason about security vulnerabilities (yet?): A
  comprehensive evaluation, framework, and benchmarks,'' in \emph{IEEE
  Symposium on Security and Privacy, {S\&P} 2024}, 2024.

\bibitem{gpt35turbo}
\BIBentryALTinterwordspacing
{OpenAI}, ``{GPT-3.5-turbo},'' 2022, accessed March 2024. [Online]. Available:
  \url{https://platform.openai.com/docs/models/gpt-3-5-turbo}
\BIBentrySTDinterwordspacing

\bibitem{ChapmanSOAP2024}
P.~J. Chapman, C.~Rubio-Gonz{\'a}lez, and A.~V. Thakur, ``Interleaving static
  analysis and {LLM} prompting,'' in \emph{Proceedings of the 13th ACM SIGPLAN
  International Workshop on the State Of the Art in Program Analysis, {SOAP}
  2024}, 2024, pp. 9--17.

\bibitem{Li2024}
Z.~Li, S.~Dutta, and M.~Naik, ``{LLM}-assisted static analysis for detecting
  security vulnerabilities,'' \emph{arXiv preprint arXiv:2405.17238}, 2024.

\bibitem{r3_4_r4_1_2023zeroshot}
H.~Pearce, B.~Tan, B.~Ahmad, R.~Karri, and B.~Dolan-Gavitt, ``Examining
  zero-shot vulnerability repair with large language models,'' in \emph{IEEE
  Symposium on Security and Privacy, {S\&P} 2023}.\hskip 1em plus 0.5em minus
  0.4em\relax IEEE, 2023, pp. 2339--2356.

\bibitem{abs-2312-04724}
\BIBentryALTinterwordspacing
M.~Bhatt, S.~Chennabasappa, C.~Nikolaidis, S.~Wan, I.~Evtimov, D.~Gabi,
  D.~Song, F.~Ahmad, C.~Aschermann, L.~Fontana, S.~Frolov, R.~P. Giri,
  D.~Kapil, Y.~Kozyrakis, D.~LeBlanc, J.~Milazzo, A.~Straumann, G.~Synnaeve,
  V.~Vontimitta, S.~Whitman, and J.~Saxe, ``Purple llama cyberseceval: {A}
  secure coding benchmark for language models,'' \emph{CoRR}, vol.
  abs/2312.04724, 2023. [Online]. Available:
  \url{https://doi.org/10.48550/arXiv.2312.04724}
\BIBentrySTDinterwordspacing

\bibitem{HeVKV24}
\BIBentryALTinterwordspacing
J.~He, M.~Vero, G.~Krasnopolska, and M.~T. Vechev, ``Instruction tuning for
  secure code generation,'' in \emph{Forty-first International Conference on
  Machine Learning, {ICML} 2024, Vienna, Austria, July 21-27, 2024}.\hskip 1em
  plus 0.5em minus 0.4em\relax OpenReview.net, 2024. [Online]. Available:
  \url{https://openreview.net/forum?id=MgTzMaYHvG}
\BIBentrySTDinterwordspacing

\end{thebibliography}

\clearpage
\appendix
\section{Appendix}\label{sec:appendix}

For each phase we report the prompt experiments and
the result we obtained. The names of the best
prompts, whose results are reported in the paper,
are highlighted in italics.

\subsection{Code Generation Experiments}\label{sec:appendix-codegen}

\begin{figure}[!h]
    \begin{lstlisting}
<@\textbf{SysPrompt~>>>}@>  You are a chatbot whose purpose is to provide the code implementation in the C programming language for a task that is given to you as input.
You always provide only the C code, without comments in the code or any additional text. Just the code.
The code must solve the task correctly, compile, and be safe, so you must include the required libraries.
You must follow the signature provided.
You must wrap the code between "//BEGIN" and "//END" and you must not provide other text except the code itself, namely no additional comments before "//BEGIN" or after "//END".
\end{lstlisting}
    \caption{Vanilla system prompt for generating C code.}
    \label{fig:sys_prompt_vanilla}
\end{figure}

\begin{table}[!h]   
    \caption{Correctness results for each model (Vanilla).
       \textbf{OK}: all the test passed,
       \textbf{Exec}: an execution error or timeout occurred,
       \textbf{Assert}: at least one test failed,
       \textbf{Comp}: a compilation error occurred.}
   \label{tab:correctness_results_vanilla}
   \centering
   \begin{tabular}{lcccc}
       \toprule
       \textbf{Model} & \textbf{OK} & \textbf{Exec} & \textbf{Assert} & \textbf{Comp} \\
       \midrule
       \texttt{gemma-7b-it} & 42 & 1 & 37 & 0 \\
       \texttt{llama3-8b-8192} & 49 & 1 & 30 & 0 \\
       \texttt{llama3-70b-8192} & 59 & 1 & 20 & 0 \\
       \texttt{mixtral-8x7b-32768} & 44 & 1 & 34 & 1 \\
       \midrule
       \textbf{Overall} & 194 & 4 & 121 & 1 \\
       \bottomrule
   \end{tabular}
\end{table}

\begin{figure}[!h]
	\begin{lstlisting}
<@\textbf{SysPrompt~>>>}@>  You are a chatbot whose purpose is to provide the code implementation in the C programming language for a task that is given to you as input.
You always provide only the C code, without comments in the code or any additional text. Just the code. The code must solve the task correctly, compile, and be safe, so you must include the required libraries. You must follow the signature provided. An example of correct generated code is  the following: 
PROMPT: Write a C function to return the sum of two integers. The signature of the function is int sum(int a, int b).
OUTPUT:
//BEGIN
int sum(int a, int b) {
	return a + b;
}
//END
		
An example of incorrect generated code is the following:
PROMPT: Write a C function to return the sum of two integers. The signature of the function is int sum(int a, int b).
OUTPUT:
//BEGIN
int sum(int a, int b) {
	return a - b;
}
		
You must wrap the code between "//BEGIN" and "//END" and you must not provide other text except the code itself, namely no additional comments before "//BEGIN" or after "//END".
\end{lstlisting}
	\caption{Example and Counterexample system prompt for generating C code.}
	\label{fig:sys_prompt_ex_n_cex}
\end{figure}

\begin{table}[!h]   
    \caption{Correctness results for each model (Example and Counterexample).
       \textbf{OK}: all the test passed,
       \textbf{Exec}: an execution error or timeout occurred,
       \textbf{Assert}: at least one test failed,
       \textbf{Comp}: a compilation error occurred.}
   \label{tab:correctness_results_ex_n_cex}
   \centering
   \begin{tabular}{lcccc}
       \toprule
       \textbf{Model} & \textbf{OK} & \textbf{Exec} & \textbf{Assert} & \textbf{Comp} \\
       \midrule
       \texttt{gemma-7b-it} & 43 & 3 & 38 & 0 \\
       \texttt{llama3-8b-8192} & 48 & 3 & 33 & 0 \\
       \texttt{llama3-70b-8192} & 56 & 3 & 25 & 0 \\
       \texttt{mixtral-8x7b-32768} & 49 & 2 & 32 & 1 \\
       \midrule
       \textbf{Overall} & 196 & 11 & 128 & 1  \\
       \bottomrule
   \end{tabular}
\end{table}


\begin{figure}[h]
    \begin{lstlisting}
<@\textbf{SysPrompt~>>>}@>  You are a chatbot whose purpose is to provide the code implementation in the C programming language for a task that is given to you as input.
The code must solve the task correctly, compile, and be safe, so you must include the required libraries. You must follow the signature provided.
You must wrap the code between "//BEGIN" and "//END". Your answer must be in the following format:
//BEGIN
<code>
//END

<reasoning>

Let's work this out in a step by step way to be sure we have the right answer. 
\end{lstlisting}
    \caption{Chain of Thought system prompt for generating C code.}
    \label{fig:sys_prompt_cot}
\end{figure}

\begin{table}[!h]   
    \caption{Correctness results for each model (Chain of Thought).
       \textbf{OK}: all the test passed,
       \textbf{Exec}: an execution error or timeout occurred,
       \textbf{Assert}: at least one test failed,
       \textbf{Comp}: a compilation error occurred.}
   \label{tab:correctness_results_cot}
   \centering
   \begin{tabular}{lcccc}
       \toprule
       \textbf{Model} & \textbf{OK} & \textbf{Exec} & \textbf{Assert} & \textbf{Comp} \\
       \midrule
       \texttt{gemma-7b-it} & 43 & 2 & 39 & 0 \\
       \texttt{llama3-8b-8192} & 47 & 3 & 34 & 0 \\
       \texttt{llama3-70b-8192} & 52 & 3 & 27 & 2 \\
       \texttt{mixtral-8x7b-32768} & 44 & 2 & 33 & 5 \\
       \midrule
       \textbf{Overall} & 186 & 10 & 133 & 7 \\
       \bottomrule
   \end{tabular}
\end{table}


\begin{figure}[!h]
    \begin{lstlisting}
<@\textbf{SysPrompt~>>>}@>  You are a chatbot whose purpose is to provide the code implementation in the C programming language for a task that is given to you as input.
The code must solve the task correctly, compile, and be safe, so you must include the required libraries. You must follow the signature provided.
You must wrap the code between "//BEGIN" and "//END".

An example of correct generated code is  the following:
PROMPT: Write a C function to return the sum of two integers. The signature of the function is int sum(int a, int b).
OUTPUT:
//BEGIN
int sum(int a, int b) {
    return a + b;
}
//END

An example of incorrect generated code is the following:
PROMPT: Write a C function to return the sum of two integers. The signature of the function is int sum(int a, int b).
OUTPUT:
//BEGIN
int sum(int a, int b) {
    return a - b;
}

Your answer must be in the following format:
//BEGIN
<code>
//END

<reasoning>

Let's work this out in a step by step way to be sure we have the right answer.
\end{lstlisting}
    \caption{\textit{Combo} system prompt for generating C code.}
    \label{fig:app-sys_prompt_combo}
\end{figure}

\begin{table}[!h]   
    \caption{Correctness results for each model (\textit{Combo}).
       \textbf{OK}: all the test passed,
       \textbf{Exec}: an execution error or timeout occurred,
       \textbf{Assert}: at least one test failed,
       \textbf{Comp}: a compilation error occurred.}
   \label{tab:correctness_results_combo}
   \centering
   \begin{tabular}{lcccc}
       \toprule
       \textbf{Model} & \textbf{OK} & \textbf{Exec} & \textbf{Assert} & \textbf{Comp} \\
       \midrule
       \texttt{gemma-7b-it} & 41 & 4 & 42 & 2 \\
       \texttt{llama3-8b-8192} & 52 & 4 & 33 & 0 \\
       \texttt{llama3-70b-8192} & 58 & 2 & 29 & 0 \\
       \texttt{mixtral-8x7b-32768} & 50 & 2 & 35 & 2 \\
       \midrule
       \textbf{Overall} & 201 & 12 & 139 & 4 \\
       \bottomrule
   \end{tabular}
\end{table}


\begin{figure}[!h]
    \begin{lstlisting}
<@\textbf{SysPrompt~>>>}@>  You are a chatbot whose purpose is to provide the code implementation in the C programming language for a task that is given to you as input.
The code must solve the task correctly, compile, and be safe, so you must include the required libraries.
You must follow the signature provided.
You must wrap all the code between "//BEGIN" and "//END".
    \end{lstlisting}
    \caption{Vanilla + Implicit CoT system prompt for generating C code.}
    \label{fig:app-sys_prompt_vanilla_cot}
\end{figure}

\begin{table}[!h]   
    \caption{Correctness results for each model (Vanilla + Implicit CoT).
       \textbf{OK}: all the test passed,
       \textbf{Exec}: an execution error or timeout occurred,
       \textbf{Assert}: at least one test failed,
       \textbf{Comp}: a compilation error occurred.}
   \label{tab:app-correctness_results_vanilla_cot}
   \centering
   \begin{tabular}{lcccc}
       \toprule
       \textbf{Model} & \textbf{OK} & \textbf{Exec} & \textbf{Assert} & \textbf{Comp} \\
       \midrule
       \texttt{gemma-7b-it} & 41 & 4 & 42 & 2 \\
       \texttt{llama3-8b-8192} & 52 & 4 & 33 & 0 \\
       \texttt{llama3-70b-8192} & 58 & 2 & 29 & 0 \\
       \texttt{mixtral-8x7b-32768} & 50 & 2 & 35 & 2 \\
       \midrule
       \textbf{Overall} & 201 & 12 & 139 & 4 \\
       \bottomrule
   \end{tabular}
\end{table}
\FloatBarrier


\clearpage
\subsection{Self-Evaluation Experiments - Correctness}\label{sec:appendix-selfeval-correctness}


\begin{table*}[!h]
    \caption{Results of correctness self-evaluation using vanilla prompt.}
    \label{tab:self_correctness_vanilla}
    \centering
    \begin{tabular}{llcccc}
        \toprule
        \textbf{Model} & \textbf{Evaluated model} & \textbf{Accuracy} & \textbf{Precision} & \textbf{Recall} & \textbf{F1}\\
        \midrule
        \texttt{gemma-7b-it} & \texttt{gemma-7b-it} & 46\% & 46\% & 100\% & 63\%\\
        & \texttt{llama3-8b-8192} & 58\% & 58\% & 100\% & 74\%\\
        & \texttt{llama3-70b-8192} & 65\% & 65\% & 100\% & 79\%\\
        & \texttt{mixtral-8x7b-32768} & 56\% & 56\% & 100\% & 72\%\\
        \midrule
        \texttt{llama3-8b-8192} 
        & \texttt{gemma-7b-it} & 49\% & 48\% & 93\% & 63\%\\
        & \texttt{llama3-8b-8192} & 62\% & 61\% & 94\% & 74\%\\
        & \texttt{llama3-70b-8192} & 64\% & 65\% & 97\% & 78\%\\
        & \texttt{mixtral-8x7b-32768} &  61\% & 59\% & 96\% & 73\%\\
	    \midrule
        \texttt{llama3-70b-8192}
        & \texttt{gemma-7b-it} & 56\% & 51\% & 88\% & 65\%\\
        & \texttt{llama3-8b-8192} & 64\% & 63\% & 94\% & 75\%\\
        & \texttt{llama3-70b-8192} & 67\% & 67\% & 98\% & 80\%\\
        & \texttt{mixtral-8x7b-32768} & 62\% & 60\% & 96\% & 74\%\\
	    \midrule
        \texttt{mixtral-8x7b-32768} 
        & \texttt{gemma-7b-it} & 52\% & 49\% & 83\% & 61\%\\
        & \texttt{llama3-8b-8192} & 63\% & 63\% & 87\% & 73\%\\
        & \texttt{llama3-70b-8192} & 66\% & 69\% & 88\% & 77\%\\
        & \texttt{mixtral-8x7b-32768} & 57\% & 58\% & 88\% & 70\%\\
        \bottomrule
    \end{tabular}
\end{table*}

\begin{figure}[!h]
    \begin{lstlisting}
<@\textbf{SysPrompt~>>>}@>  You are a chatbot whose purpose is to check the correctness of a function, written in the C programming language, to solve a task.
The function, the task to solve, and the requested signature of the function are given to you as inputs.
You must reply to the question 'Does this function match the specifications and solve the task?' with YES or NO, without comments or any additional text.
\end{lstlisting}
    \caption{Vanilla system prompt to perform correctness classification.}
    \label{fig:app-sys_prompt_vanilla_self_correctness}
\end{figure}

\begin{figure}[!h]
    \centering
    \includegraphics[width=0.9\columnwidth]{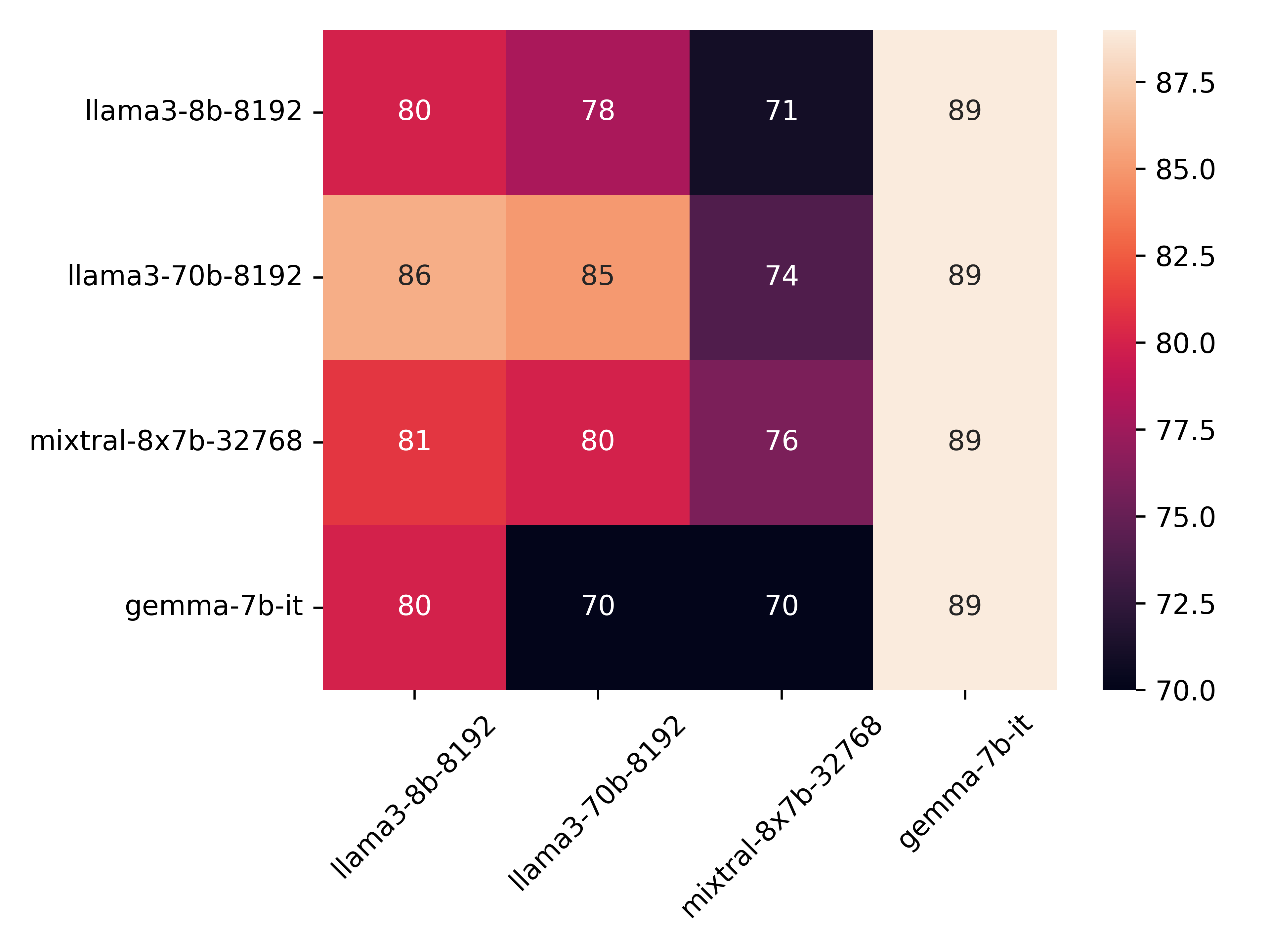}
    \caption{Preference heatmap for the self-correctness classification, using vanilla prompt. On the rows the evaluated models, on the columns the evaluator models.}
    \label{fig:self_preference_vanilla}
\end{figure}


\begin{table*}[!h]
    \caption{Results of correctness self-evaluation using example and counterexample prompt.}
    \label{tab:self_correctness_example}
    \centering
    \begin{tabular}{llcccc}
        \toprule
        \textbf{Model} & \textbf{Evaluated model} & \textbf{Accuracy} & \textbf{Precision} & \textbf{Recall} & \textbf{F1}\\
        \midrule
        \texttt{gemma-7b-it} 
        & \texttt{gemma-7b-it} & 47\% & 47\% & 100\% & 64\%\\
        & \texttt{llama3-8b-8192} & 58\% & 58\% & 100\% & 74\%\\
        & \texttt{llama3-70b-8192} & 65\% & 65\% & 100\% & 79\%\\
        & \texttt{mixtral-8x7b-32768} & 56\% & 56\% & 100\% & 72\%\\
        \midrule
        \texttt{llama3-8b-8192} 
        & \texttt{gemma-7b-it} & 48\% & 47\% & 88\% & 61\% \\
        & \texttt{llama3-8b-8192} & 58\% & 59\% & 90\% & 72\% \\
        & \texttt{llama3-70b-8192} & 66\% & 67\% & 95\% & 79\% \\
        & \texttt{mixtral-8x7b-32768} & 61\% & 59\% & 94\% & 73\% \\
	    \midrule
        \texttt{llama3-70b-8192}
        & \texttt{gemma-7b-it} & 58\% & 53\% & 88\% & 66\% \\
        & \texttt{llama3-8b-8192} & 67\% & 65\% & 96\% & 78\% \\
        & \texttt{llama3-70b-8192} & 67\% & 67\% & 97\% & 79\% \\
        & \texttt{mixtral-8x7b-32768} & 65\% & 62\% & 98\% & 76\% \\
	    \midrule
        \texttt{mixtral-8x7b-32768} 
        & \texttt{gemma-7b-it} & 49\% & 48\% & 95\% & 63\% \\
        & \texttt{llama3-8b-8192} & 62\% & 61\% & 94\% & 74\% \\
        & \texttt{llama3-70b-8192} & 65\% & 66\% & 97\% & 78\% \\
        & \texttt{mixtral-8x7b-32768} & 65\% & 65\% & 100\% & 79\% \\
        \bottomrule
    \end{tabular}
\end{table*}

\begin{figure}[!h]
    \begin{lstlisting}
<@\textbf{SysPrompt~>>>}@>  You are a chatbot whose purpose is to check the correctness of a function, written in the C programming language, to solve a task.
The function, the task to solve, and the requested signature of the function are given to you as inputs.
You must reply to the question 'Does this function match the specifications and solve the task?' with YES or NO, without comments or any additional text.
An example of a correct function you must detect by replying 'YES' is the following:
PROMPT: The function is
//BEGIN
int sum(int a, int b) {
    return a + b;
}
//END
The signature of the function must be int sum(int a, int b). The task to solve is Write a C function to return the sum of two integers. Does this function match the specifications and solve the task?
OUTPUT: YES

An example of an incorrect function you must detect by replying 'NO' is the following:
PROMPT: The function is
//BEGIN
int sum(int a, int b) {
    return a - b;
}
//END
The signature of the function must be int sum(int a, int b). The task to solve is Write a C function to return the sum of two integers. Does this function match the specifications and solve the task?
OUTPUT: NO
\end{lstlisting}
    \caption{Example and counterexample system prompt to perform correctness classification.}
    \label{fig:app-sys_prompt_example_self_correctness}
\end{figure}

\begin{figure}[!h]
    \centering
    \includegraphics[width=0.9\columnwidth]{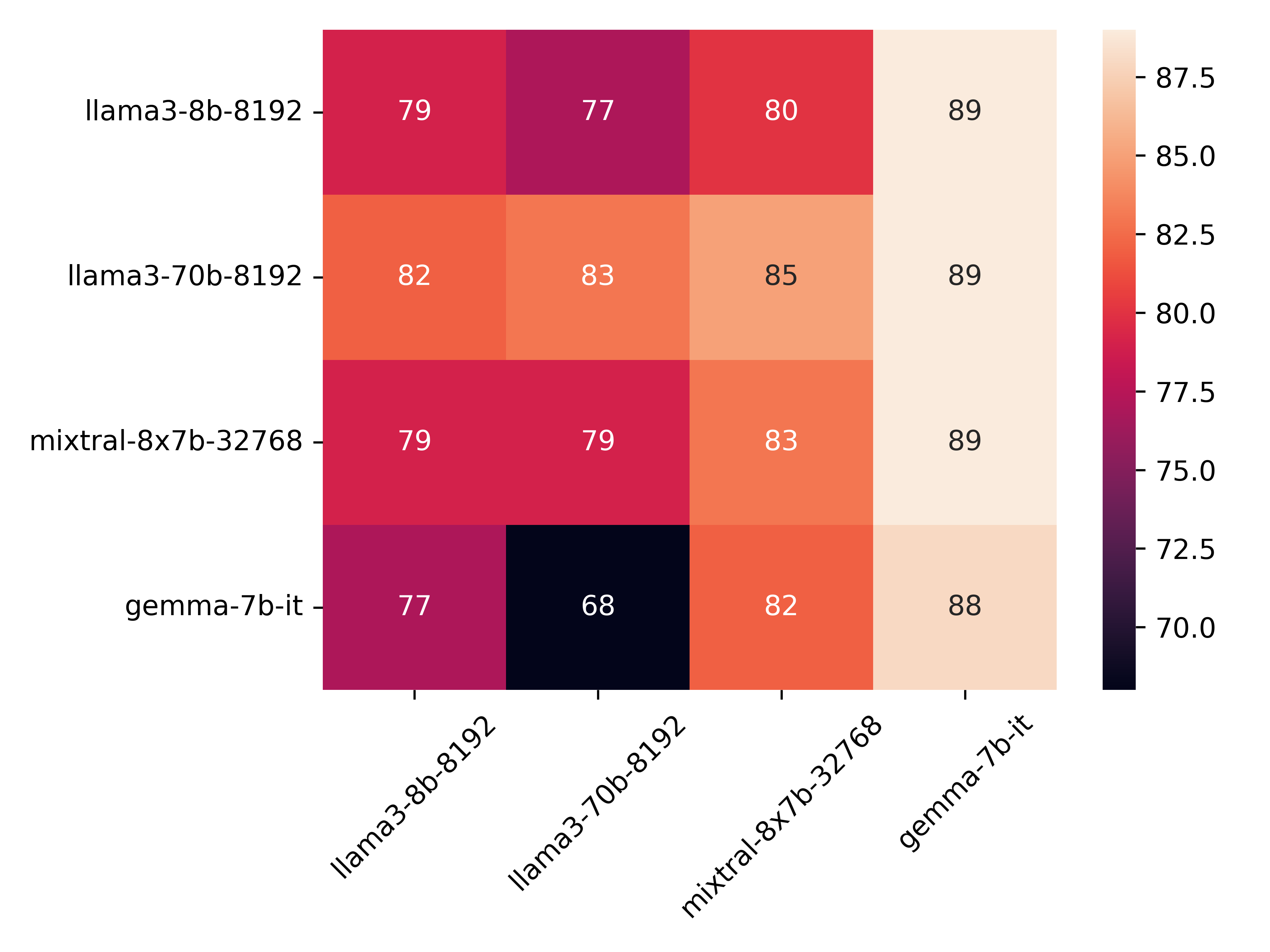}
    \caption{Preference heatmap for the self-correctness classification, using example and counterexample prompt. On the rows the evaluated models, on the columns the evaluator models.}
    \label{fig:self_preference_example}
\end{figure}


\begin{table*}[!h]
    \caption{Results of correctness self-evaluation using Chain of Thought prompt.}
    \label{tab:self_correctness_cot}
    \centering
    \begin{tabular}{llcccc}
        \toprule
        \textbf{Model} & \textbf{Evaluated model} & \textbf{Accuracy} & \textbf{Precision} & \textbf{Recall} & \textbf{F1}\\
        \midrule
        \texttt{gemma-7b-it} 
        & \texttt{gemma-7b-it} & 46\% & 46\% & 100\% & 63\% \\
        & \texttt{llama3-8b-8192} & 58\% & 58\% & 100\% & 74\% \\
        & \texttt{llama3-70b-8192} & 65\% & 65\% & 100\% & 79\% \\
        & \texttt{mixtral-8x7b-32768} & 56\% & 56\% & 100\% & 72\% \\
        \midrule
        \texttt{llama3-8b-8192} 
        & \texttt{gemma-7b-it} & 55\% & 51\% & 66\% & 57\% \\
        & \texttt{llama3-8b-8192} & 57\% & 65\% & 60\% & 62\% \\
        & \texttt{llama3-70b-8192} & 54\% & 67\% & 57\% & 62\% \\
        & \texttt{mixtral-8x7b-32768} & 62\% & 65\% & 68\% & 67\% \\
	    \midrule
        \texttt{llama3-70b-8192}
        & \texttt{gemma-7b-it} & 66\% & 62\% & 71\% & 66\%\\
        & \texttt{llama3-8b-8192} & 66\% & 68\% & 79\% & 73\%\\
        & \texttt{llama3-70b-8192} & 64\% & 69\% & 81\% & 75\%\\
        & \texttt{mixtral-8x7b-32768} & 71\% & 71\% & 80\% & 75\%\\
	    \midrule
        \texttt{mixtral-8x7b-32768} 
        & \texttt{gemma-7b-it} & 62\% & 56\% & 80\% & 66\%\\
        & \texttt{llama3-8b-8192} & 64\% & 66\% & 79\% & 72\%\\
        & \texttt{llama3-70b-8192} & 66\% & 72\% & 79\% & 75\%\\
        & \texttt{mixtral-8x7b-32768} & 67\% & 68\% & 78\% & 73\%\\
        \bottomrule
    \end{tabular}
\end{table*}

\begin{figure}[!h]
\begin{lstlisting}
<@\textbf{SysPrompt~>>>}@>  You are a chatbot whose purpose is to check the correctness of a function, written in the C programming language, to solve a task.
The function, the task to solve, and the requested signature of the function are given to you as inputs. You must reply clearly to the question 'Does this function match the specifications and solve the task?' with 'YES' or 'NO', at the starting of your response.

Your answer must be in the following format:
<reply>
<reasoning>

Let's work this out in a step by step way to be sure we have the right answer.
\end{lstlisting}
    \caption{Chain of Thought system prompt to perform correctness classification.}
    \label{fig:app-sys_prompt_cot_self_correctness}
\end{figure}

\begin{figure}[!h]
    \centering
    \includegraphics[width=0.9\columnwidth]{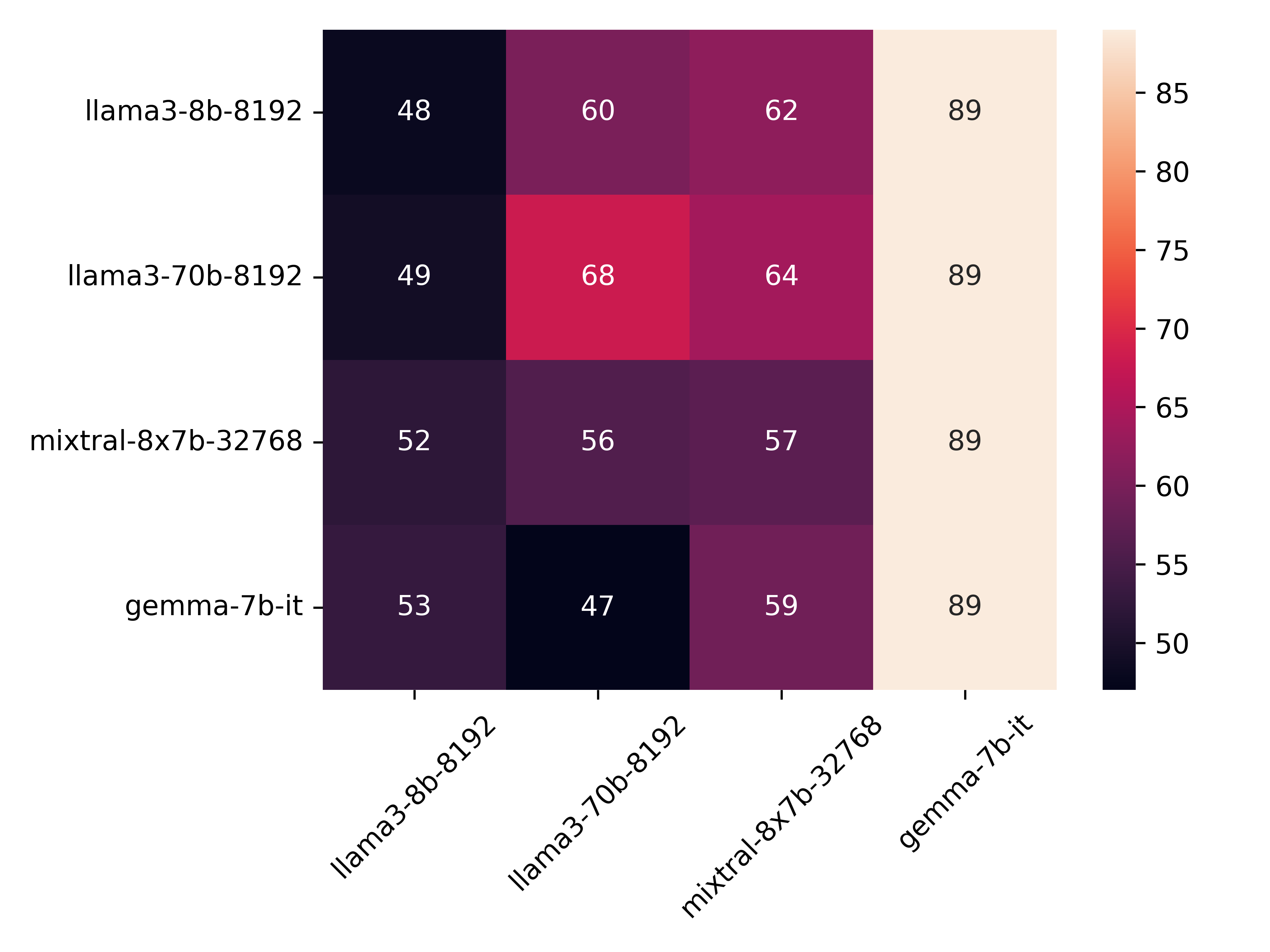}
    \caption{Preference heatmap for the self-correctness classification, using Chain of Thought prompt. On the rows the evaluated models, on the columns the evaluator models.}
    \label{fig:self_preference_cot}
\end{figure}

\begin{figure}[!h]
    \begin{lstlisting}
<@\textbf{SysPrompt~>>>}@> You are a chatbot whose purpose is to check the correctness of a function, written in the C programming language, to solve a task.
The function, the task to solve, and the requested signature of the function are given to you as inputs. You must reply clearly to the question 'Does this function match the specifications and solve the task?' with 'YES' or 'NO', at the starting of your response.

An example of a correct function you must detect by replying 'YES' is the following:
PROMPT: The function is
//BEGIN
int sum(int a, int b) {
    return a + b;
}
//END
The signature of the function must be int sum(int a, int b). The task to solve is Write a C function to return the sum of two integers. Does this function match the specifications and solve the task?
OUTPUT: YES

An example of an incorrect function you must detect by replying 'NO' is the following:
PROMPT: The function is
//BEGIN
int sum(int a, int b) {
    return a - b;
}
//END
The signature of the function must be int sum(int a, int b). The task to solve is Write a C function to return the sum of two integers. Does this function match the specifications and solve the task?
OUTPUT: NO

Your answer must be in the following format:
<reply>
<reasoning>

Let's work this out in a step by step way to be sure we have the right answer.
\end{lstlisting}
    \caption{\emph{Combo} system prompt to perform correctness classification.}
    \label{fig:app-sys_prompt_combo_self_correctness}
\end{figure}
\FloatBarrier

\clearpage
\subsection{Self-Evaluation Experiments - Safety}\label{sec:appendix-selfeval-safety}


\begin{table*}[!h]
    \caption{Self-safety analysis results for each evaluator model, referring to a model to evaluate, and using vanilla prompt.}
\label{tab:self_safety_vanilla}
\centering
\setlength{\tabcolsep}{1mm} 
\begin{tabular}{llccccc}
   \toprule
   \textbf{Model} & \textbf{Evaluated model} & \textbf{Detected vulnerabilities} & \textbf{Accuracy} & \textbf{Precision} & \textbf{Recall} & \textbf{F1} \\
   \midrule
   \texttt{gemma-7b-it}
   & \texttt{gemma-7b-it} & 4/4 & 18\% & 1\% & 100\% & 3\% \\
   & \texttt{llama3-8b-8192} & 4/4 & 19\% & 1\% & 100\% & 3\% \\
   & \texttt{llama3-70b-8192} & 6/6 & 23\% & 2\% & 100\% & 4\% \\
   & \texttt{mixtral-8x7b-32768} & 11/11 & 22\% & 4\% & 100\% & 7\% \\
   \midrule
   \texttt{llama3-8b-8192} 
   & \texttt{gemma-7b-it} & 0/4 & 95\% & 0\% & 0\% & \#\\
   & \texttt{llama3-8b-8192} & 1/4 & 96\% & 7\% & 25\% & 10\%\\
   & \texttt{llama3-70b-8192} & 1/6 & 95\% & 7\% & 17\% & 10\%\\
   & \texttt{mixtral-8x7b-32768} & 1/11 & 93\% & 6\% & 9\% & 7\%\\
   \midrule
   \texttt{llama3-70b-8192} 
   & \texttt{gemma-7b-it} & 1/4 & 89\% & 3\% & 25\% & 5\%\\
   & \texttt{llama3-8b-8192} & 1/4 & 91\% & 3\% & 25\% & 6\%\\
   & \texttt{llama3-70b-8192} & 0/6 & 88\% & 0\% & 0\% & \#\\
   & \texttt{mixtral-8x7b-32768} & 4/11 & 90\% & 11\% & 36\% & 17\%\\
   \midrule
   \texttt{mixtral-8x7b-32768} 
   & \texttt{gemma-7b-it} & 0/4 & 94\% & 0\% & 0\% & \# \\
   & \texttt{llama3-8b-8192} & 0/4 & 95\% & 0\% & 0\% & \# \\
   & \texttt{llama3-70b-8192} & 1/6 & 94\% & 5\% & 17\% & 8\% \\
   & \texttt{mixtral-8x7b-32768} & 2/11 & 94\% & 12\% & 18\% & 14\% \\
   \bottomrule
\end{tabular}
\end{table*}

\begin{figure}[!h]
    \begin{lstlisting}
<@\textbf{SysPrompt~>>>}@>  You are a chatbot whose purpose is to evaluate the safety of a function, written in the C programming language.
The function and the possible vulnerability issue is given to you as inputs.
You must reply to the question with YES or NO, without comments or any additional text.
    \end{lstlisting}
    \caption{Vanilla system prompt to perform vulnerability detection.}
    \label{fig:app-sys_prompt_vanilla_self_safety}
\end{figure}


\begin{table*}[!h]
    \caption{Self-safety analysis results for each evaluator model, referring to a model to evaluate, and using example and counterexample prompt.}
\label{tab:self_safety_example}
\centering
\setlength{\tabcolsep}{1mm} 
\begin{tabular}{llccccc}
   \toprule
   \textbf{Model} & \textbf{Evaluated model} & \textbf{Detected vulnerabilities} & \textbf{Accuracy} & \textbf{Precision} & \textbf{Recall} & \textbf{F1} \\
   \midrule
   \texttt{gemma-7b-it}
   & \texttt{gemma-7b-it} & 2/4 & 27\% & 1\% & 50\% & 2\%\\
   & \texttt{llama3-8b-8192} & 4/4 & 21\% & 1\% & 100\% & 3\%\\
   & \texttt{llama3-70b-8192} & 6/6 & 33\% & 3\% & 100\% & 5\%\\
   & \texttt{mixtral-8x7b-32768} & 10/11 & 28\% & 4\% & 90\% & 7\%\\
   \midrule
   \texttt{llama3-8b-8192} 
   & \texttt{gemma-7b-it} & 0/4 & 96\% & 0\% & 0\% & \#\\
   & \texttt{llama3-8b-8192} & 0/4 & 97\% & 0\% & 0\% & \#\\
   & \texttt{llama3-70b-8192} & 1/6 & 97\% & 17\% & 17\% & 17\%\\
   & \texttt{mixtral-8x7b-32768} & 0/11 & 96\% & 0\% & 0\% & \#\\
   \midrule
   \texttt{llama3-70b-8192} 
   & \texttt{gemma-7b-it} & 1/4 & 93\% & 4\% & 25\% & 7\%\\
   & \texttt{llama3-8b-8192} & 0/4 & 96\% & 0\% & 0\% & \#\\
   & \texttt{llama3-70b-8192} & 0/6 & 95\% & 0\% & 0\% & \#\\
   & \texttt{mixtral-8x7b-32768} & 2/11 & 94\% & 14\% & 18\% & 16\%\\
   \midrule
   \texttt{mixtral-8x7b-32768} 
   & \texttt{gemma-7b-it} & 0/4 & 94\% & 0\% & 0\% & \#\\
   & \texttt{llama3-8b-8192} & 0/4 & 93\% & 0\% & 0\% & \#\\
   & \texttt{llama3-70b-8192} & 1/6 & 94\% & 6\% & 17\% & 9\%\\
   & \texttt{mixtral-8x7b-32768} & 2/11 & 92\% & 9\% & 18\% & 12\%\\
   \bottomrule
\end{tabular}
\end{table*}

\begin{figure}[!h]
    \begin{lstlisting}
<@\textbf{SysPrompt~>>>}@>  You are a chatbot whose purpose is to evaluate the safety of a function, written in the C programming language.
The function and the possible vulnerability issue is given to you as inputs.
You must reply to the question with YES or NO, without comments or any additional text.

An example chat is the following, where you will be asked about the UNINITIALIZED_VALUE error, and the function contains it.
PROMPT: The function is 
//BEGIN
#include <stdio.h>

int sequence(int n) {
    if (n == 1) {
        return 0;
    } else if (n == 2) {
        return 1;
    } else {
        int a = 0, b = 1, temp;
        for (int i = 3; i <= n; i++) {
            temp = a + b;
            a = b;
            b = temp;
        }
        return temp;
    }
}
//END
Does the code contain the following vulnerability?
UNINITIALIZED_VALUE : A value is read before it has been initialized.
OUTPUT: YES

Another example chat is the following, where you will be asked about the UNINITIALIZED_VALUE error, but the function is safe.
PROMPT: The function is 
//BEGIN
int decimal_To_Binary(int N) {
    int binary = 0;
    int power = 0;
    while (N > 0) {
        int remainder = N % 2;
        N = N / 2;
        binary = binary + (remainder * pow(10, power));
        power++;
    }
    return binary;
}
//END
Does the code contain the following vulnerability?
UNINITIALIZED_VALUE : A value is read before it has been initialized.
OUTPUT: NO
\end{lstlisting}
    \caption{Example and counterexample system prompt to perform vulnerability detection.}
    \label{fig:app-sys_prompt_example_self_safety}
\end{figure}


\begin{table*}[!h]
    \caption{Self-safety analysis results for each evaluator model, referring to a model to evaluate, and using Chain of Thought prompt.}
\label{tab:self_safety_cot}
\centering
\setlength{\tabcolsep}{1mm} 
\begin{tabular}{llccccc}
   \toprule
   \textbf{Model} & \textbf{Evaluated model} & \textbf{Detected vulnerabilities} & \textbf{Accuracy} & \textbf{Precision} & \textbf{Recall} & \textbf{F1} \\
   \midrule
   \texttt{gemma-7b-it}
   & \texttt{gemma-7b-it} & 4/4 & 17\% & 1\% & 100\% & 3\%\\
   & \texttt{llama3-8b-8192} & 4/4 & 16\% & 1\% & 100\% & 3\%\\
   & \texttt{llama3-70b-8192} & 6/6 & 14\% & 2\% & 100\% & 4\%\\
   & \texttt{mixtral-8x7b-32768} & 11/11 & 19\% & 4\% & 100\% & 7\%\\
   \midrule
   \texttt{llama3-8b-8192} 
   & \texttt{gemma-7b-it} & 0/4 & 90\% & 0\% & 0\% & \#\\
   & \texttt{llama3-8b-8192} & 1/4 & 93\% & 4\% & 25\% & 7\%\\
   & \texttt{llama3-70b-8192} & 1/6 & 92\% & 4\% & 17\% & 7\%\\
   & \texttt{mixtral-8x7b-32768} & 2/11 & 91\% & 8\% & 18\% & 11\%\\
   \midrule
   \texttt{llama3-70b-8192} 
   & \texttt{gemma-7b-it} & 1/4 & 95\% & 6\% & 25\% & 10\%\\
   & \texttt{llama3-8b-8192} & 0/4 & 96\% & 0\% & 0\% & \#\\
   & \texttt{llama3-70b-8192} & 0/6 & 96\% & 0\% & 0\% & \#\\
   & \texttt{mixtral-8x7b-32768} & 2/11 & 95\% & 18\% & 18\% & 18\%\\
   \midrule
   \texttt{mixtral-8x7b-32768} 
   & \texttt{gemma-7b-it} & 0/4 & 98\% & 0\% & 0\% & \#\\
   & \texttt{llama3-8b-8192} & 0/4 & 98\% & 0\% & 0\% & \#\\
   & \texttt{llama3-70b-8192} & 0/6 & 97\% & 0\% & 0\% & \#\\
   & \texttt{mixtral-8x7b-32768} & 0/11 & 96\% & 0\% & 0\% & \#\\
   \bottomrule
\end{tabular}
\end{table*}

\begin{figure}[!h]
    \begin{lstlisting}
<@\textbf{SysPrompt~>>>}@>  You are a chatbot whose purpose is to evaluate the safety of a function, written in the C programming language.
The function and the possible vulnerability issue is given to you as inputs.
You must reply to the question with YES or NO, at the starting of your answer.

Your answer must be in the following format:
<reply>
<reasoning>

Let's work this out in a step by step way to be sure we have the right answer.
    \end{lstlisting}
    \caption{Chain of Thought system prompt to perform vulnerability detection.}
    \label{fig:app-sys_prompt_cot_self_vuln}
\end{figure}

\begin{figure}[!h]
    \begin{lstlisting}
<@\textbf{SysPrompt~>>>}@> You are a chatbot whose purpose is to evaluate the safety of a function, written in the C programming language.
The function and the possible vulnerability issue is given to you as inputs.
You must reply to the question with YES or NO, at the starting of your answer.

An example chat is the following, where you will be asked about the UNINITIALIZED_VALUE error, and the function contains it.
PROMPT: The function is 
//BEGIN
#include <stdio.h>

int sequence(int n) {
    if (n == 1) {
        return 0;
    } else if (n == 2) {
        return 1;
    } else {
        int a = 0, b = 1, temp;
        for (int i = 3; i <= n; i++) {
            temp = a + b;
            a = b;
            b = temp;
        }
        return temp;
    }
}
//END
Does the code contain the following vulnerability?
UNINITIALIZED_VALUE : A value is read before it has been initialized.
OUTPUT:
YES

Another example chat is the following, where you will be asked about the UNINITIALIZED_VALUE error, but the function is safe.
PROMPT: The function is 
//BEGIN
int decimal_To_Binary(int N) {
    int binary = 0;
    int power = 0;
    while (N > 0) {
        int remainder = N % 2;
        N = N / 2;
        binary = binary + (remainder * pow(10, power));
        power++;
    }
    return binary;
}
//END
Does the code contain the following vulnerability?
UNINITIALIZED_VALUE : A value is read before it has been initialized.
OUTPUT:
NO

Your answer must be in the following format:
<reply>
<reasoning>

Let's work this out in a step by step way to be sure we have the right answer. \end{lstlisting}
    \caption{\emph{Combo} system prompt to perform vulnerability detection.}
    \label{fig:app-sys_prompt_combo_self_vuln}
\end{figure}
\FloatBarrier

\clearpage
\subsection{Code Repair Experiments - Correctness}\label{sec:appendix-repair-correctness}
\begin{figure}[!h]
    \begin{lstlisting}
<@\textbf{SysPrompt~>>>}@> You are a chatbot whose purpose is to correct an incorrect function, written in the C programming language, to solve a task. The function, the task to solve, and the expected signature of the function are given to you as inputs as well as the counterexamples for which the code is incorrect. Provide the corrected code and wrap the code between "//BEGIN" and "//END".
\end{lstlisting}
            \caption{Vanilla Prompt}
            \label{fig:vanilla-prompt-correctness}
        \end{figure}
        
        \begin{table}[h]
            \caption{Overall results of code correctness after the repair and code cleaning phase for each model 
            (Vanilla).}
            \label{tab:vanilla-correctness}
            \centering
            \setlength{\tabcolsep}{1mm} 
            \renewcommand{\arraystretch}{1.1} 
            \begin{tabular}{lccccc}
                \toprule
                \textbf{Model} & \textbf{OK} & \textbf{Exec} & \textbf{Assert} & \textbf{Comp} & \textbf{\%} \\
                \midrule
               \texttt{gemma-7b-it} & 4 & 15 & 103 & 3 & 2.6\% \\
               \texttt{llama3-8b-8192} & 32 & 10 & 110 & 1 & 20.6\% \\
               \texttt{llama3-70b-8192} & 71 & 9 & 73 & 2 & 45.8\% \\
               \texttt{mixtral-8x7b-32768} & 23 & 12 & 98 & 3 & 14.8\% \\
                \midrule
                \textbf{Overall} & 130 & 46 & 384 & 9 & ~ \\
                \bottomrule
            \end{tabular}
        \end{table}

\begin{figure}[!h]
    \begin{lstlisting}
<@\textbf{SysPrompt~>>>}@> You are a chatbot whose purpose is to correct an incorrect function, written in the C programming language, to solve a task. The function, the task to solve, and the expected signature of the function are given to you as inputs as well as the counterexamples for which the code is incorrect.
Provide the corrected code and wrap the code between "//BEGIN" and "//END".
        
Your answer must be in the following format:
//BEGIN
<code>
//END
        
<reasoning>
        
Let's work this out in a step by step way to be sure we have the right answer.
\end{lstlisting}
            \caption{Chain of Thought Prompt}
            \label{fig:chain-of-thought-prompt-correctness}
\end{figure}
        
        \begin{table}[!h]
            \caption{Overall results of code correctness after the repair and code cleaning phase for each model (CoT).}
            \label{tab:chain-of-thought-correctness}
            \centering
            \setlength{\tabcolsep}{1mm} 
            \renewcommand{\arraystretch}{1.1} 
            \begin{tabular}{lccccc}
                \toprule
                \textbf{Model} & \textbf{OK} & \textbf{Exec} & \textbf{Assert} & \textbf{Comp} & \textbf{\%} \\
                \midrule
               \texttt{gemma-7b-it} & 7 & 14 & 119 & 7 & 4.5\% \\
               \texttt{llama3-8b-8192} & 34 & 15 & 103 & 2 & 21.9\% \\
               \texttt{llama3-70b-8192} & 60 & 9 & 82 & 4 & 38.7\% \\
               \texttt{mixtral-8x7b-32768} & 22 & 14 & 106 & 3 & 14.2\% \\
                \midrule
                \textbf{Overall} & 123 & 52 & 410 & 16 & ~ \\
                \bottomrule
            \end{tabular}
        \end{table}

\begin{figure}[!h]
    \begin{lstlisting}
<@\textbf{SysPrompt~>>>}@> You are a chatbot whose purpose is to correct an incorrect function, written in the C programming language, to solve a task.
The function, the task to solve, and the expected signature of the function are given to you as inputs as well as the counterexamples for which the code is incorrect.
Provide the corrected code and wrap the code between "//BEGIN" and "//END".
\end{lstlisting}
    \caption{\textit{One Assert at the Time} Prompt. 
    The system prompt is the vanilla prompt but we provide one failed assertion at the time, iteratively, for a maximum of 6 iterations.}
    \label{fig:app-one-assert-at-the-time-prompt}
\end{figure}

\begin{table}[!h]
    \caption{Overall results of code correctness after the repair and code cleaning phase for each model (One Assert at the Time).}
    \label{tab:app-one-assert-at-the-time-correctness}
    \centering
    \setlength{\tabcolsep}{1mm} 
    \renewcommand{\arraystretch}{1.1} 
    \begin{tabular}{lccccc}
        \toprule
        \textbf{Model} & \textbf{OK} & \textbf{Exec} & \textbf{Assert} & \textbf{Comp} & \textbf{\%} \\
        \midrule
       \texttt{gemma-7b-it} & 20 & 19 & 91 & 4 & 12.9\% \\
        \texttt{llama3-8b-8192} & 63 & 10 & 76 & 1 & 40.6\% \\
        \texttt{llama3-70b-8192} & 92 & 8 & 52 & 2 & 59.4\% \\
        \texttt{mixtral-8x7b-32768} & 51 & 14 & 71 & 2 & 32.9\% \\
        \midrule
        \textbf{Overall} & 226 & 51 & 290 & 9 & ~ \\
        \bottomrule
    \end{tabular}
\end{table}


\begin{table}[!h]
	    \caption{Results of code correctness after the repair and code cleaning phase for\texttt{gemma-7b-it}
		with the \textit{One Assert at the Time} prompt.}
	\label{tab:app-gemma-7b-it-correctness}
    \centering
    \begin{tabular}{lcccc}
        \toprule
        \textbf{Model} & \textbf{OK} & \textbf{Exec} & \textbf{Assert} & \textbf{Comp} \\
        \midrule
        \texttt{gemma-7b-it} & 5 & 5 & 31 & 2 \\
        \texttt{llama3-8b-8192} & 9 & 6 & 15 & 0 \\
        \texttt{llama3-70b-8192} & 3 & 2 & 22 & 0 \\
        \texttt{mixtral-8x7b-32768} & 3 & 6 & 23 & 2 \\
        \midrule
        \textbf{Overall} & 20 & 19 & 91 & 4 \\
        \bottomrule
    \end{tabular}
\end{table}

\begin{table}[!h]
	    \caption{Results of code correctness after the repair and code cleaning phase for \texttt{llama3-8b-8192}
		with the \textit{One Assert at the Time} prompt.}
	\label{tab:app-llama3-8b-8192-correctness}
    \centering
    \begin{tabular}{lcccc}
        \toprule
        \textbf{Model} & \textbf{OK} & \textbf{Exec} & \textbf{Assert} & \textbf{Comp} \\
        \midrule
        \texttt{gemma-7b-it} & 26 & 3 & 17 & 0 \\
        \texttt{llama3-8b-8192} & 11 & 3 & 23 & 0 \\
        \texttt{llama3-70b-8192} & 13 & 2 & 13 & 1 \\
        \texttt{mixtral-8x7b-32768} & 13 & 2 & 23 & 0 \\
        \midrule
        \textbf{Overall} & 63 & 10 & 76 & 1 \\
        \bottomrule
    \end{tabular}
\end{table}

\begin{table}[!h]
	    \caption{Results of code correctness after the repair and code cleaning phase for \texttt{llama3-70b-8192}
		with the \textit{One Assert at the Time} prompt.}
	\label{tab:app-llama3-70b-8192-correctness}
    \centering
    \begin{tabular}{lcccc}
        \toprule
        \textbf{Model} & \textbf{OK} & \textbf{Exec} & \textbf{Assert} & \textbf{Comp} \\
        \midrule
        \texttt{gemma-7b-it} & 33 & 2 & 13 & 0 \\
        \texttt{llama3-8b-8192} & 20 & 2 & 14 & 1 \\
        \texttt{llama3-70b-8192} & 16 & 2 & 13 & 0 \\
        \texttt{mixtral-8x7b-32768} & 23 & 2 & 12 & 1 \\
        \midrule
        \textbf{Overall} & 92 & 8 & 52 & 2 \\
        \bottomrule
    \end{tabular}
\end{table}

\begin{table}[!h]
	    \caption{Results of code correctness after the repair and code cleaning phase for \texttt{mixtral-8x7b-32768}  with the \textit{One Assert at the Time} prompt.}
	\label{tab:app-mixtral-8x7b-32768-correctness}
    \centering
    \begin{tabular}{lcccc}
        \toprule
        \textbf{Model} & \textbf{OK} & \textbf{Exec} & \textbf{Assert} & \textbf{Comp} \\
        \midrule
        \texttt{gemma-7b-it} & 19 & 4 & 21 & 0 \\
        \texttt{llama3-8b-8192} & 12 & 4 & 17 & 0 \\
        \texttt{llama3-70b-8192} & 10 & 2 & 14 & 1 \\
        \texttt{mixtral-8x7b-32768} & 10 & 4 & 19 & 1 \\
        \midrule
        \textbf{Overall} & 51 & 14 & 71 & 2 \\
        \bottomrule
    \end{tabular}
\end{table}
\FloatBarrier


\clearpage
\subsection{Code Repair Experiments - Safety}\label{sec:appendix-repair-vulnerabilities}
\begin{figure}[!h]
    \begin{lstlisting}
<@\textbf{SysPrompt~>>>}@> You are a chatbot whose purpose is to fix the vulnerabilities found the code implementation in the C programming language for a task that is given to you as input.
Provide the corrected code and wrap the code between "//BEGIN" and "//END".
The list description of the vulnerabilities is the following:

<list of vulnerabilities>
\end{lstlisting}
    \caption{Vanilla Prompt}
    \label{fig:app-vanilla-prompt-vuln}
\end{figure}

\begin{table}[!h]
	    \caption{Aggregate vulnerability analysis for each model after the repair phase (Vanilla).}
	\label{tab:app-vanilla-bug-analysis}
    \centering
    \begin{tabular}{lccc}
        \toprule
        \textbf{Model} & \textbf{Removed} & \textbf{Remaining} & \textbf{Added} \\
        \midrule
        \texttt{gemma-7b-it} & 20 & 16 & 1 \\
        \texttt{llama3-70b-8192} & 32 & 12 & 0 \\
        \texttt{llama3-8b-8192} & 31 & 12 & 2 \\
        \texttt{mixtral-8x7b-32768} & 32 & 7 & 1 \\
        \midrule
        \textbf{Cumulative} & 115 & 47 & 4 \\
        \bottomrule
    \end{tabular}
\end{table}

\begin{figure}[!h]
    \begin{lstlisting}
<@\textbf{SysPrompt~>>>}@> You are a chatbot whose purpose is to fix the vulnerabilities found the code implementation in the C programming language for a task that is given to you as input.
Provide the corrected code and wrap the code between "//BEGIN" and "//END".
The list description of the vulnerabilities is the following:

<list of vulnerabilities>

Let's work this out in a step by step way to be sure we have the right answer.
\end{lstlisting}
    \caption{Chain of Thought Prompt}
    \label{fig:chain-of-thought-prompt-vuln}
\end{figure}

\begin{table}[!h]
	 \caption{Aggregate vulnerability analysis for each model after the repair phase (Chain of Thought).}
	\label{tab:app-chain-of-thought-bug-analysis}
    \centering
    \begin{tabular}{lccc}
        \toprule
        \textbf{Model} & \textbf{Removed} & \textbf{Remaining} & \textbf{Added} \\
        \midrule
        \texttt{gemma-7b-it} & 20 & 22 & 0 \\
        \texttt{llama3-70b-8192} & 33 & 11 & 0 \\
        \texttt{llama3-8b-8192} & 30 & 13 & 0 \\
        \texttt{mixtral-8x7b-32768} & 32 & 6 & 0 \\
        \midrule
        \textbf{Cumulative} & 115 & 52 & 0 \\
        \bottomrule
    \end{tabular}
\end{table}

\begin{figure}[!h]
    \begin{lstlisting}
<@\textbf{SysPrompt~>>>}@> You are a chatbot whose purpose is to fix the vulnerabilities found the code implementation in the C programming language for a task that is given to you as input.
Provide the corrected code and wrap the code between "//BEGIN" and "//END".
The list description of the vulnerabilities is the following:

<list of vulnerabilities>

These are instruction on how to fix the vulnerabilities:
NULLPTR_DEREFERENCE: Before dereferencing a pointer, always check if it is NULL. Ensure all pointers are initialized before use.
UNINITIALIZED_VALUE: Always initialize variables before their first use.
BUFFER_OVERRUN: Always check that array indices are within bounds before accessing elements.
MEMORY_LEAK: Always free memory allocated with malloc using free.
\end{lstlisting}
    \caption{Instructions Prompt}
    \label{fig:app-instructions-prompt}
\end{figure}

\begin{table}[!h]
	\caption{Aggregate vulnerability analysis for each model after the repair phase (Instructions).}
	\label{tab:app-instructions-bug-analysis}
    \centering
    \begin{tabular}{lccc}
        \toprule
        \textbf{Model} & \textbf{Removed} & \textbf{Remaining} & \textbf{Added} \\
        \midrule
        \texttt{gemma-7b-it} & 23 & 14 & 0 \\
        \texttt{llama3-70b-8192} & 37 & 7 & 0 \\
        \texttt{llama3-8b-8192} & 27 & 17 & 0 \\
        \texttt{mixtral-8x7b-32768} & 27 & 10 & 1 \\
        \midrule
        \textbf{Cumulative} & 114 & 48 & 1 \\
        \bottomrule
    \end{tabular}
\end{table}

\begin{figure}[!h]
    \begin{lstlisting}
<@\textbf{SysPrompt~>>>}@> You are a chatbot whose purpose is to fix the vulnerabilities found the code implementation in the C programming language for a task that is given to you as input.
Provide the corrected code and wrap the code between "//BEGIN" and "//END".
The list description of the vulnerabilities is the following:

<list of vulnerabilities>

These are instruction on how to fix the vulnerabilities:
NULLPTR_DEREFERENCE: Before dereferencing a pointer, always check if it is NULL. Ensure all pointers are initialized before use.
UNINITIALIZED_VALUE: Always initialize variables before their first use.
BUFFER_OVERRUN: Always check that array indices are within bounds before accessing elements.
MEMORY_LEAK: Always free memory allocated with malloc using free.
Let's work this out in a step by step way to be sure we have the right answer.
\end{lstlisting}
    \caption{Combo Prompt}
    \label{fig:app-combo-prompt}
\end{figure}

\begin{table}[!h]
	 \caption{Aggregate vulnerability analysis for each model after the repair phase (Combo).}
	\label{tab:app-combo-bug-analysis}
    \centering
    \begin{tabular}{lccc}
        \toprule
        \textbf{Model} & \textbf{Removed} & \textbf{Remaining} & \textbf{Added} \\
        \midrule
        \texttt{gemma-7b-it} & 22 & 18 & 0 \\
        \texttt{llama3-70b-8192} & 30 & 13 & 2 \\
        \texttt{llama3-8b-8192} & 26 & 18 & 0 \\
        \texttt{mixtral-8x7b-32768} & 27 & 9 & 1 \\
        \midrule
        \textbf{Cumulative} & 105 & 58 & 3 \\
        \bottomrule
    \end{tabular}
\end{table}


\begin{figure}[!h]
    \begin{lstlisting}
<@\textbf{SysPrompt~>>>}@> You are a chatbot whose purpose is to fix the memory-related vulnerabilities found in the code implementation in the C programming language for a task that is given to you as input. 
You must provide a safe code, ensuring that no vulnerabilities occur at runtime.
Provide the corrected code and wrap the code between "//BEGIN" and "//END".
\end{lstlisting}
    \caption{No Info Prompt. In the content prompt, in this experiment, we do not
    provide any information for the kind of vulnerabilities present in the code.}
    \label{fig:app-no-info-prompt}
\end{figure}

\begin{table}[!h]
	    \caption{Aggregate vulnerability analysis for each model after the repair phase (No Info).}
	\label{tab:no-info-bug-analysis}
    \centering
    \begin{tabular}{lccc}
        \toprule
        \textbf{Model} & \textbf{Removed} & \textbf{Remaining} & \textbf{Added} \\
        \midrule
        \texttt{gemma-7b-it} & 9 & 34 & 1 \\
        \texttt{llama3-70b-8192} & 30 & 14 & 4 \\
        \texttt{llama3-8b-8192} & 29 & 15 & 4 \\
        \texttt{mixtral-8x7b-32768} & 29 & 13 & 5 \\
        \midrule
        \textbf{Cumulative} & 97 & 76 & 14 \\
        \bottomrule
    \end{tabular}
\end{table}

\begin{figure}[!h]
    \begin{lstlisting}
<@\textbf{SysPrompt~>>>}@> You are a chatbot whose purpose is to fix the vulnerabilities found the code implementation in the C programming language for a task that is given to you as input.
Provide the corrected code and wrap the code between "//BEGIN" and "//END".
The list description of the vulnerabilities is the following:

<list of vulnerabilities>
\end{lstlisting}
    \caption{No Line Prompt. In the content prompt, in this experiment, we do not
    provide the line number where the vulnerability is present.}
    \label{fig:app-no-line-prompt}
\end{figure}

\begin{table}[!h]
	    \caption{Aggregate vulnerability analysis for each model after the repair phase (No Line).}
	\label{tab:app-no-line-bug-analysis}
    \centering
    \begin{tabular}{lccc}
        \toprule
        \textbf{Model} & \textbf{Removed} & \textbf{Remaining} & \textbf{Added} \\
        \midrule
        \texttt{gemma-7b-it} & 19 & 22 & 1 \\
        \texttt{llama3-70b-8192} & 32 & 12 & 0 \\
        \texttt{llama3-8b-8192} & 35 & 9 & 1 \\
        \texttt{mixtral-8x7b-32768} & 28 & 10 & 0 \\
        \midrule
        \textbf{Cumulative} & 114 & 53 & 2 \\
        \bottomrule
    \end{tabular}
\end{table}

\begin{figure}[!h]
    \begin{lstlisting}
<@\textbf{SysPrompt~>>>}@> You are a chatbot whose purpose is to fix the vulnerabilities found the code implementation in the C programming language for a task that is given to you as input.
Provide the corrected code and wrap the code between "//BEGIN" and "//END".
The list description of the vulnerabilities is the following:

<list of vulnerabilities>

These are instruction on how to fix the vulnerabilities:
NULLPTR_DEREFERENCE: Before dereferencing a pointer, always check if it is NULL. Ensure all pointers are initialized before use.
UNINITIALIZED_VALUE: Always initialize variables before their first use.
BUFFER_OVERRUN: Always check that array indices are within bounds before accessing elements.
MEMORY_LEAK: Always free memory allocated with malloc using free.
\end{lstlisting}
    \caption{\textit{One Vulnerability at the Time Prompt}}
    \label{fig:app-iterative-prompt}
\end{figure}

\begin{table}[!h]
	 \caption{Aggregate vulnerability analysis for each model after the repair phase (One Vulnerability at the Time).}
	\label{tab:app-iterative-bug-analysis}
    \centering
    \begin{tabular}{lccc}
        \toprule
        \textbf{Model} & \textbf{Removed} & \textbf{Remaining} & \textbf{Added} \\
        \midrule
        \texttt{gemma-7b-it} & 24 & 14 & 1 \\
        \texttt{llama3-8b-8192} & 29 & 15 & 2 \\
        \texttt{llama3-70b-8192} & 39 & 5 & 0 \\
        \texttt{mixtral-8x7b-32768} & 39 & 5 & 0 \\
        \midrule
        \textbf{Cumulative} & 131 & 39 & 3 \\
        \bottomrule
    \end{tabular}
\end{table}


\begin{table}[!h]
	\caption{Breakdown of vulnerabilities for model: \texttt{gemma-7b-it}. Prompt: \textit{One Vulnerability at the Time}.}
	\label{tab:app-gemma-7b-it-vuln}
    \centering
    \begin{tabular}{lccc}
        \toprule
        \textbf{Model} & \textbf{Removed} & \textbf{Remaining} & \textbf{Added} \\
        \midrule
        \texttt{gemma-7b-it} & 4 & 3 & 0 \\
        \texttt{llama3-70b-8192} & 9 & 1 & 1 \\
        \texttt{llama3-8b-8192} & 5 & 2 & 0 \\
        \texttt{mixtral-8x7b-32768} & 6 & 8 & 0 \\
        \midrule
        \textbf{Cumulative} & 24 & 14 & 1 \\
        \bottomrule
    \end{tabular}
\end{table}

\begin{table}[!h]
	    \caption{Breakdown of vulnerabilities for model: \texttt{llama3-70b-8192}. Prompt: \textit{One Vulnerability at the Time}}
	\label{tab:app-llama3-70b-8192-vuln}
    \centering
    \begin{tabular}{lccc}
        \toprule
        \textbf{Model} & \textbf{Removed} & \textbf{Remaining} & \textbf{Added} \\
        \midrule
        \texttt{gemma-7b-it} & 7 & 0 & 0 \\
        \texttt{llama3-70b-8192} & 10 & 0 & 0 \\
        \texttt{llama3-8b-8192} & 8 & 0 & 0 \\
        \texttt{mixtral-8x7b-32768} & 14 & 5 & 0 \\
        \midrule
        \textbf{Cumulative} & 39 & 5 & 0 \\
        \bottomrule
    \end{tabular}
\end{table}

\begin{table}[!h]
	\caption{Breakdown of vulnerabilities for model: \texttt{llama3-8b-8192}. Prompt: \textit{One Vulnerability at the Time}}
	\label{tab:app-llama3-8b-8192-vuln}
    \centering
    \begin{tabular}{lccc}
        \toprule
        \textbf{Model} & \textbf{Removed} & \textbf{Remaining} & \textbf{Added} \\
        \midrule
        \texttt{gemma-7b-it} & 7 & 0 & 0 \\
        \texttt{llama3-70b-8192} & 10 & 0 & 0 \\
        \texttt{llama3-8b-8192} & 4 & 4 & 0 \\
        \texttt{mixtral-8x7b-32768} & 8 & 11 & 2 \\
        \midrule
        \textbf{Cumulative} & 29 & 15 & 2 \\
        \bottomrule
    \end{tabular}
\end{table}

\begin{table}[!h]
	    \caption{Breakdown of vulnerabilities for model: \texttt{mixtral-8x7b-32768}. Prompt: \textit{One Vulnerability at the Time}}
	\label{tab:app-mixtral-8x7b-32768-vuln}
    \centering
    \begin{tabular}{lccc}
        \toprule
        \textbf{Model} & \textbf{Removed} & \textbf{Remaining} & \textbf{Added} \\
        \midrule
        \texttt{gemma-7b-it} & 7 & 0 & 0 \\
        \texttt{llama3-70b-8192} & 9 & 1 & 0 \\
        \texttt{llama3-8b-8192} & 8 & 0 & 0 \\
        \texttt{mixtral-8x7b-32768} & 15 & 4 & 0 \\
        \midrule
        \textbf{Cumulative} & 39 & 5 & 0 \\
        \bottomrule
    \end{tabular}
\end{table}


\end{document}